\documentclass[11pt]{article}

\usepackage{geometry, amsmath, amsthm, latexsym, amssymb, graphicx}
\usepackage{algorithm}

\usepackage{natbib}

\usepackage{subfigure}

\usepackage{hyperref}
\hypersetup{citecolor=black,colorlinks=false}

\newcommand{\beq}{\begin{equation}}
\newcommand{\eeq}{\end{equation}}
\newcommand{\ber}{\begin{eqnarray}}
\newcommand{\eer}{\end{eqnarray}}

\newcommand{\xb}{\boldsymbol{x}}

\newcommand{\bs}{\boldsymbol}

\newcommand{\bit}{\begin{itemize}}
\newcommand{\eit}{\end{itemize}}

\newcommand{\ben}{\begin{enumerate}}
\newcommand{\een}{\end{enumerate}}

\numberwithin{equation}{section}

\newcommand{\V}{\boldsymbol{V}}

\title{Bayesian Uncertainty Quantification of Local Volatility Model}
\author{ Kai Yin \ and Anirban Mondal \\
Department of Mathematics, Applied Mathematics, and Statistics \\
Case Western Reserve University, Cleveland, OH 44106 USA  }

\begin{document}

\maketitle

\begin{abstract}
Local volatility is an important quantity in option pricing, portfolio hedging, and risk management. It is not directly observable from the market; hence calibrations of local volatility models are necessary using observed market data. Unlike most existing point-estimate methods, we cast the large-scale nonlinear inverse problem into the Bayesian framework, yielding a posterior distribution of the local volatility, which naturally quantifies its uncertainty. This extra uncertainty information enables traders and risk managers to make better decisions. To alleviate the computational cost, we apply Karhunen--L\`oeve expansion to reduce the dimensionality of the Gaussian Process prior for local volatility. A modified two-stage adaptive Metropolis algorithm is applied to sample the posterior probability distribution, which further reduces computational burdens caused by repetitive numerical forward option pricing model solver and time of heuristic tuning. We demonstrate our methodology with both synthetic and market data.   
\end{abstract}


\section{Introduction}

Since the introduction of the seminal Black-Scholes (B-S) model \citep{black1973pricing} for European options, many complex volatility models have been developed to capture underlying asset dynamics in different financial markets. Such models need to be calibrated to the corresponding observable market data before they can be applied for financial derivative pricing and hedging. European option prices are commonly used for model calibrations due to their high liquidity. The calibration process is critical because inaccurate pricing can result in a high risk of introducing arbitrage into the market.

In the B-S model, volatility is assumed to be a constant, which does not explain the generic volatility smile and smirk. Various research efforts have been dedicated to relaxing this constant volatility assumption to match the empirical market observation. There are mainly two types of established methods. The first is stochastic approaches such as Generalized AutoRegressive Conditional Heteroskedasticity (GARCH) model \citep{bollerslev1988capital}, stochastic volatility models \citep{Hull1987ThePO} with or without jump-diffusion \citep{andersen2000jump}, and the second is deterministic approaches like local volatility models \citep{dupire1994pricing}). Among these models, the local volatility model maintains the market completeness like in the B-S model, which is important in the sense of allowing arbitrage-free pricing and hedging \citep{dupire1994pricing}. Comparing to stochastic volatility models, the local volatility model can reproduce any given market European option data; hence, the estimated local volatility is capable to be used for exotic option pricing and hedging \citep{shreve2004stochastic} without losing consistency to the market \citep{dupire1994pricing}.

In this article, we focus on calibrating the local volatility model using European option price data. Such local volatility estimation is a large-scale nonlinear ill-posed inverse problem, in which the unknown is a function of time and spot prices and the data size is limited. The forward option pricing model is nonlinear, and the market data is corrupted by noises, which is only observable up to bid and ask. These features create the ill-posedness of the inverse problem. Many researchers have tried to address these problems for the last decades. In general, they can be classified into three categories. The first one is to apply Dupire's formula to obtain the local volatility function after interpolating and extrapolating discrete market option data \citep{andersen1998equity, kahale2004arbitrage, fengler2009arbitrage, glaser2012arbitrage}. The drawback of this approach is that it introduces non-market prices into the calibration process, which might cause risk exposure of arbitrage. The second approach proposed by \cite{Avellaneda1997CalibratingVS} is through solving a relative-entropy minimization method to reconstruct the local volatility with assumptions of lower and upper bounds. The local volatility surface constructed by this method often has abnormal peaks and troughs close to expiration dates, which is not practical for exotic pricing purposes. The third approach is to solve this problem by optimization assuming either a parametric or nonparametric form of local volatility, such that the obtained parameters can reproduce the market option function with the smallest discrepancy with respect to some chosen measure \citep{lagnado1997technique, jackson1998computation, Coleman1999ReconstructingTU, crepey2003calibration, cezaro2008convex, geng2014non}. \citep{lagnado1997technique,crepey2003calibration, geng2014non} addressed the under-determined problem by introducing a Tikhonov regularization, which led to a well-posed optimization problem. The stability, uniqueness, and convergence of the Tikhonov regularization approach were theoretically justified in \citep{lishang2001identifying, crepey2003calibration, Egger2005TikhonovRA, hein2005some}. However, the ``best" solution might not actually be unique in the sense that the repriced options are between the range of bid and ask, so a probabilistic approach is more proper to describe the uncertainty of the unknown local volatility surfaces. 

We consider a Bayesian approach to this large-scale nonlinear inverse problem in which the unknown quantity is the local volatility surface, where the solution to the problem is a posterior probability distribution given the observed option data here. The Bayesian approach has a natural form of regularization through the prior probability distribution of the unknowns. The posterior distribution can be used to quantify uncertainties of the local volatility learned from the data. The posterior distribution is not analytically tractable as the likelihood contains the nonlinear forward option pricing model. Markov chain Monte Carlo (MCMC) based sampling methods are generally used to draw inference for Bayesian models. \cite{Tegner2019APA} used a Bayesian method with the Gaussian process (GP) prior for the local volatility surface. \cite{AnimokuUY18} have also used a Bayesian method with a constant variance and uncorrelated prior model for the volatility surface. Their prior mean is in the form of quadratic polynomials or additive splines. A GP on the other hand is fully specified by its mean and correlation structure. GP can be considered as a prior over the functional space \citep{rasmussen2003gaussian}. By considering a correlation structure over the prior it becomes more flexible in estimating the unknown form of the volatility surface than the polynomial and spline based mean surface models as used in \cite{AnimokuUY18}. However, in both \cite{Tegner2019APA} and \cite{AnimokuUY18}, the dimension of the posterior probability distribution is related to the discretization grids of the numerical forward model solver, which could be very high for achieving higher order of accuracy. This could lead to a significant computational challenge in the MCMC method. In this article we also use a GP prior for the volatility surface but address the high-dimensional issue and the corresponding computational challenges using a functional data analysis approach. In particular, we apply Karhunen Lo`eve (K-L) expansion for dimension reduction of the GP prior. Using K-L expansion the posterior dimension is reduced significantly and is no more tied to the the discretization grid of the forward solver. 

Based on the common smoothness assumption for the local volatility surface \citep{Coleman1999ReconstructingTU, geng2014non, Tegner2019APA}, here we use a square exponential correlation kernel. Before K-L expansion can be practically applied to dimension reduction for a stochastic process, a Fredholm equation of the second kind must be solved to obtain its corresponding eigenvalues and vectors. Its corresponding K-L expansion is usually achieved numerically since it has no analytical expression with respect to the Lebesgue measure. Thus, for an MCMC-based method, the Fredholm equation of second kind \citep{le2010spectral} must be solved numerically and repetitively for each iteration, which is as expensive as a full Gaussian Process inference. We address this issue by considering an analytical K-L decomposition of the square exponential kernel based GP process with respect to Gaussian measure \citep{zhu1997gaussian, rasmussen2003gaussian}. This allows a functional representation of the local volatility function in the reduced dimension space. Moreover, using a separable covariance kernel, the analytical K-L eigenpairs in two dimensions (time and asset price) can be obtained by the tensor product of the corresponding eigenpairs in each dimension. 

The forward option pricing model has no closed-form solution; hence it is usually solved by either Monte Carlo simulation of the expected value of discounted option pay-off function with respect to risk-neutral measure or numerical solver for the corresponding Partial Differential Equations (PDE). Here we develop a finite element method (FEM) to solve Dupire’s PDE following \cite{hirsa2012computational}, which is numerically more stable than the traditional finite difference method for irregular and coarse discretization. As options are heavily traded near at-the-money with shorter maturities, and option prices in that area are more sensitive to volatility, the use of irregular discretization is very important for the numerical accuracy of the solver. Using the FEM, solution can be obtained at any grid as it is based upon a piece-wise representation of the solution in terms of specified basis functions. Hence no post-interpolation is necessary to match the solution of the forward model with the observed option data in an irregular grid. The use of a stable coarse discretization also facilitates an efficient Monte Carlo sampling algorithm as described in the next paragraph..

Although the K-L expansion is applied for dimension reduction, the remaining number of parameters to be estimated is still substantial. The MCMC-based sampling algorithm is still computationally expensive for repetitive computations of the forward simulation and high dimensionality in the posterior distribution. We use a two-stage adaptive Metropolis Algorithm (TSAM) \cite{Mondal2021ATS}. In the first stage of the proposed algorithm, an adaptive proposal is used based on the previously sampled states and the corresponding acceptance probability is computed based on an approximated likelihood where the forward model is solved in a coarse grid. The true expensive likelihood is evaluated while computing the second stage acceptance probability only if the proposal is accepted in the first stage. The adaptive nature of the algorithm guarantees faster convergence of the chain and very good mixing properties. On the other hand, the two-stage approach helps in rejecting the bad proposals in the inexpensive first stage, making the algorithm computationally efficient. Our methodology is first demonstrated using simulated data sets with the local volatility generated from a known one-dimension function, a known two-dimension function, and finally a known GP in two-dimension. The numerical results show that the proposed method can adequately predict the benchmark local volatility surfaces. Then the proposed method is applied to real market option data from S\&P 500 index and Euro Stoxx 50 index. The underlying local volatility is unknown for real data, thus the predicted option prices are compared with the real market option data, and predictions are well in the 95\% credible bands. 

The article is structured as follows: Section 2 briefly introduces the local volatility model and the set-up of its corresponding inverse problem; Section 3 casts the large scale nonlinear inverse problem into a Bayesian framework; Section 4 contains the numerical solver for the forward option pricing model, and TSAM algorithm to sample the posterior distribution of local volatility; Section 5 includes the deployment of our methodology to synthetic data and real market option data; Section 6 concludes the paper and points out future research opportunities.

\section{Local Volatilty Model}

In this section, we briefly introduce the forward local volatility model and the corresponding inverse problem set-up. To begin with, we consider a financial market with a continuously finite trading time interval $[0,t_{max}]$, which is formed by the following two financial products: a risk-less bank account and a stock, the price process of which is modeled by the local volatility model. We further assume that there are no fractional costs, like taxes or dividends.

Local volatility model assumes that the underlying asset price process $(S_t)_{t \in [0,t_{max}]}$ follows a general diffusion equation:
\begin{equation} \label{eq::underlyingProcess}
dS_t = \mu_t S_t dt + \sigma(t, S_t) S_t dW_t,
\end{equation}
where drift $\mu_t$ is the expected rate of return, local volatility $\sigma: \mathbb{R}^+ \times [0,t_{max}] \rightarrow \mathbb{R}^+$ is a deterministic function depending on both asset price $S_t$ and time $t$, and $W_t$ is a standard Brownian motion. Local volatility is the only non-observable parameter in this model, which we try to imply from the market option price data. Since the focus of this paper is the calibration of local volatility from option data, without loss of generality, we assume the asset price process is a risk-neutral dynamic with respect to measure $Q$.

Next, we consider an European call option written on a stock assumed above with maturity $T$ and strike $K$ at a reference time $t$, which has a pay-off function $h(S_T) = max(S_T - K,0)$. 
The value of the option can be evaluated by the risk-neutral pricing formula:
\begin{equation}\label{eq::riskNeutralPF}
    V(t,S_t, T, K, \sigma) = E^{Q}[e^{-r(T-t)}h(S_T)|S_t].
 \end{equation}

Based on Feynman-Kac formula, calculating option prices from equation (\ref{eq::riskNeutralPF}) is equivalent to solving the following partial differential equation: 
\begin{equation}\label{eq::BSlv}
\frac{\partial V}{\partial t} + (r-q) S \frac{\partial V}{\partial S} + \frac{1}{2}\sigma(t, S)^2 S^2 \frac{\partial^2 V}{\partial S^2} - rV = 0,
\end{equation}
with terminal condition $V(T,S)=max(S-K,0)$. 

By changing of variables for (\ref{eq::BSlv}), Dupire \citep{dupire1994pricing} obtained a forward equation depending on maturity $T \in [0,~ T_{max}]$ and strike $K \in [K_{min}, K_{max}]$: 
\begin{equation}\label{eq::DupireE}
\frac{\partial V}{\partial T} - \frac{1}{2}\sigma^2(K,T) K^2 \frac{\partial^2 V}{\partial K^2}+(r-q) K\frac{\partial V}{\partial K} + qV= 0, 
\end{equation}
with initial conditions
$V(K,0) = max(S_0 -K, 0)$ and boundary conditions $V(K_{min},T) = e^{(-rT)}(S_0-K_{min})$ and $V(K_{max}, T) = 0$ . 

The advantage of Dupire's forward equation for our problem set-up is that it allows us to price all call options with various strikes and maturities for a given local volatility function at once. We solve \ref{eq::DupireE} with a finite element method, details of which will be discussed in section 4. 
In reality, before applying any of the above approaches to evaluate the price of a European call option, we need to know the local volatility $\sigma$ as an input, which is usually calibrated from the market observed option prices. European options are commonly used because of their liquidity.

\subsection{The Inverse Problem}
 Without loss of generality, we let the reference time $t = 0$. Assume we have a set of N market observed European option prices of an underlying asset with n maturities $T_1$, $...$, $T_n$, and $m_i$ strikes $K_{i1},~ ...,~ K_{im_i}$ for each maturity $T_i$, where $\displaystyle N = \sum_{i=1}^{n} m_i$. In the trading market, option prices are usually only observable up to bid and ask, where differences between asks and bids depend on the liquidity of the options. Denote $V_{ij}^a$ and $V_{ij}^b$ to be the bid and ask prices correspondingly. Let $\bar{V}_{ij} = (V_{ij}^a+V_{ij}^b)/2$ to be the average of bid and ask prices, and we use it as our observed market data $V_{ij}^{obs}$ from now on.

We further assume that the observed market option prices are obtained from the sum of prices computed from the local volatility model and additive noises. For finitely many observed market option prices, we assume the following observing process
\begin{equation}\label{eq::noiseModel}
V_{ij}^{obs} = V(T_i, K_{ij}, \sigma) + \epsilon_{ij}, ~i=1,..,n  \text{ and } j = 1, ..., m_i,
\end{equation} 
where $\epsilon_{ij}s$ can be treated as a combination of measurement error and model discrepancy error, and assumed to be i.i.d Normal$(0, \sigma^2_{\epsilon})$. 

More than one local volatility surface can produce option prices between the bid and ask prices for this under-determined optimization problem instead of just a single ``best fit" solution. Thus, instead of finding a single best solution that might mislead decision makings from risk managers and traders, it is essential to describe the solution with a probability measure. To this end, the Bayesian approach provides a natural way to assign a random process prior to the local volatility function and obtain a posterior distribution as a solution for this inverse problem. 

\section{The Bayesian Framework} 
\label{section::bayesianFramework}
In this section, we introduce a general Bayesian framework to infer the local volatility $\sigma$ from the option data. The Bayesian solution of the inverse problem (\ref{eq::noiseModel}) is the posterior distribution of local volatility $\sigma$ conditioned on all the option price observations $\V^{obs}$.
Using the Bayes theorem, we can express the posterior distribution as: 

\begin{equation} \label{eq::generalBayesformula}
p(\sigma, \sigma_{\epsilon}|\V^{obs}) \propto p(\V^{obs}|\sigma, \sigma_{\epsilon})p(\sigma)p(\sigma_{\epsilon}).
\end{equation}

The problem left is to specify the probability terms on the right hand side of the expression: (i) $P(\sigma)$: the prior model for the local volatility surface, where we use the Karhunen-Lo\`eve expansion to parameterize $\sigma$ for dimension reduction purpose, (ii) $P(\sigma_{\epsilon})$: the prior for the error standard deviation, (iii) $p(\V^{obs}|\sigma)$: the likelihood function obtained from (\ref{eq::noiseModel}). We provide the modeling details of all these terms in the following subsections. 

\subsection{Modeling the Prior Process}

The prior distribution for local volatility is modeled as a spatial-temporal random field. To guarantee the positiveness of volatility, we take a log transform of it. 
Let $Y = log(\sigma)$ to be a random field equipped with a probability space $(\Omega, U, P)$, where $Y(x,\omega)$ is a realization of the random field with $x=(T,K) \in D = [0,T_{max}]\times[K_{min}, K_{max}]$ and $\omega \in \Omega$. 
For indexing convention, we denote the point $(T_i, K_{ij})$ as $x_k$, where $k = \sum_{l=1}^{i-1} m_l + j$.

We model the prior distribution Y by a two dimensional Gaussian process, which is a standard prior for functionals. The GP is a stochastic process for any finite set of realizations are multivariate Gaussian, and it can be completely determined by its mean $\mu$ and covariance kernel $R$. So here, 
\begin{equation}
Y \sim GP(\mu, R).
\end{equation}

The computational cost of a full GP-based inference is in the order of $O(N^3)$ for a discretized domain with $N$ grid points, which is very expensive when finer grids are needed for accuracy in option pricing numerical solver. For the dimensional reduction, we use the Karhunen L\`oeve expansion (K-L expansion) \citep{loeve1977elementary} to get a finite-dimensional parametric representation of $Y$. First, we briefly review Mercer's theorem \citep{konig2013eigenvalue} and K-L expansion.  

A stochastic process can be expressed as a linear combination of products of corresponding eigenpairs in an optimal $L^2$ sense with random coefficients. Without loss of generality, we assume $E(Y(x,\omega)) = 0$, then the stochastic process can be expressed as:

\begin{equation}
Y({x}, \omega) =  \sum_{i} \sigma_Y \sqrt{\lambda_i}\phi_i (x) \theta_i(\omega), 
\end{equation} 

where $(\lambda_i, \phi_i)$ are the solutions of the second Fredholm integral equation: 
\begin{equation}
\label{fred}
\int_{} R (x, x^*) \phi_i (x^*)dx^* = \lambda_i \phi_i(\boldsymbol{x}),~~~ i = 1, 2,...,
\end{equation} 

and $\theta_i$'s are i.i.d N$(0,1)$ if $Y$ is Guassian. The spatio-temporal dependence of the domain D is resolved by the $L^2$ deterministic basis $\{\phi_i\}$, and the randomness is represented by the scalar random variables $\theta_i$. By Mercer's theorem, we have $R(x, x^*) = \sum_i \lambda_i \phi_i(x)\phi_i(x^*)$. In practice, it is desirable to keep the least leading order terms but still capturing most of the energy of the stochastic process Y. 
Let the eigenvalues $\lambda_i$ be ordered as $\lambda_1 \geq \lambda_2 \geq \dots$, then for a truncated K-L expansion with $N_{KL}$ terms $Y_{N_{KL}} = \sum_{i=1}^{N_{KL}} \sqrt{\lambda_i} \theta_i \phi_i$, its energy ratio is defined as follows: 
\begin{equation}\label{eq::energyRatio}
    e(N_{KL}) = \frac{\sum_{i=1}^{N_{KL}} \lambda_i}{\sum_{i=1}^{\infty} \lambda_i}
\end{equation}

The faster the decay speed of $\lambda$, the better approximation of the K-L expansion would be in the $L^2$ sense. We choose $N_{KL}$ such that the energy ratio reaches an upper bound. 

There are different choices of kernel functions for the Gaussian process, for example, exponential, squared exponential, Matern, spherical, etc. In this case, the latent local volatility is assumed to be a smooth surface and a squared exponential kernel is a common choice, although other smooth kernels can also be used. Smoothness is known as a regularization condition for function approximations with a finite number of data points \citep{tikhonov2013numerical, wahba1990spline}. The smoothness assumption is also important for the stability of the finite element method numerical solver, which we introduce in section 4. Another advantage of using a squared exponential kernel is the existence of analytical eigen-decomposition of Fredholm integral equation \ref{fred}, with respect to Gaussian measure. The analytical solution reduces a lot of computational cost in the MCMC sampling iterations since the computational cost of solving the eigenvalue problem numerically is also in the order of $O(N^3)$, the same as inverting the covariance matrix. By using an analytical truncated K-L expansion, the dimension of the posterior in our method is tied to the amount of data from the real market nor to the number of discretization grids. The other potential benefit of using GP with the square exponential kernel is that it is infinitely differential, which makes pricing exotic options and hedging more accessible by utilizing the posterior distribution of local volatility.

Next, we describe the analytical K-L expansions for one dimensional GP with a square exponential kernel with respect to Gaussian measure, followed by the analytical two-dimensional K-L expansion for a GP with a separable square exponential kernel.

\subsubsection{Analytical K-L for Squared Exponential Kernel in One-dimension}\label{subsubsec::analytical K-L1}

For the squared exponential kernel $R(x, x') = exp(-(x-x')^2/2\ell^2)$ and measure $ \mu(x) \sim$ Normal$(0,\sigma_{\mu})$, there exists analytic expressions for the eigenvalues and eigenfunctions of the following eigen problem   \citep{zhu1997gaussian, rasmussen2003gaussian}:
\begin{equation} \label{eq::1dEigenFunGaussianMeasure}
    \int exp \Big \{ -(x-x')^2/l \Big \} \phi_k(x,x') d\mu(x) = \lambda_i \phi_i(x'), ~ i = 1, 2, 3, ...
\end{equation}

The corresponding eigenvalues $\lambda_k$ and eigenfunctions $\phi_k$ are given by 
\begin{equation}\label{eq::GmeasureAnalytical}
    \lambda_k = \sqrt{\frac{2a}{A}}B^k, ~~~    \phi_k(x) =h_k exp(-(c-a)x^2)H_k(\sqrt{2c}x)
\end{equation}

where $H_k(x) = (-1)^kexp(x^2)\frac{d^k}{dx^k} exp(-x^2)$ is the $k_{th}$ order Hermite polynomial, $a^{-1} = 4\sigma^2$, $b^{-1}= 2\ell^2$ and 
\begin{equation}
    c = \sqrt{a^2 + 2ab}, ~~~ A= a+b+c, ~~~B=b/A.
\end{equation}

For a fixed number of eigenpairs, the choice of measure standard deviation $\sigma_{\mu}$ is critical to balance the rich representation of the eigenfunctions and the total percentage of variance it captures. If $\sigma_{\mu}$ goes to infinity, then the normal measure converges to a Lebesgue measure over the whole real line $\mathbb{R}$. In that scenario, we would need to keep a large number of eigenpairs to capture 90\% variances, which is equivalent to K-L with respect to the Lebesgue measure. On the other hand, if we let $\sigma_{\mu}$ go to 0, then we will capture all the variability with only one eigen-pair, which obviously limits the richness of representation of the eigenfunctions. We use a simulation study using the possible values of the correlation parameters in a grid and choose an optimal $\sigma_{\mu}$ balancing the rich representation and the variability.  For the choices of the center, it is very natural to place the measure center to the current price along the strike direction since the trading volume of options spares away for the at-the-money. Along the maturity direction, we place the measure center in the middle between the reference time and the end time we want to forecast.

\subsubsection{Analytical K-L for Squared Exponential Kernel in Two-dimension}\label{subsubsec::analytical K-L}
From the one dimensional analytical eigen decomposition (\ref{eq::GmeasureAnalytical}), we can easily obtain the analytical K-L decomposition for a Gaussian process with separable square exponential kernels $R(x,x') = \sigma_Y^2 exp \Big \{ -(x_1-x'_1)^2/l_1-(x_2-x'_2)^2/l_2 \Big \}$ in two-dimension. 

Generally, the two-dimensional eigen-decomposition can be obtained by solving the following second Fredholm equation:
\begin{equation}\label{eq::twodEigenFun}
    \int \int exp \Big \{ -(x_1-x'_1)^2/l_1-(x_2-x'_2)^2/l_2 \Big \} \phi_k(x,x') d\mu(x) = \lambda_k \phi_k(x'), ~ k = 1, 2, 3, ...
\end{equation}

Instead of solving the equation (\ref{eq::twodEigenFun}), we use the result from the one-dimensional case to come up with the analytical K-L decomposition for the two-dimensional case. In general, if the kernel is separable, then the multi-dimensional eigenpairs can be obtained from the tensor product of the corresponding one-dimension eigenpairs, and hence we have the following result.

{\bf Result 1}: Suppose $\{\lambda_{1i}, ~\phi_{1i} \}_{i=1,2,3,...}$ are eigen pairs of equation \ref{eq::1dEigenFunGaussianMeasure} along T direction, and $\{\lambda_{2j}, ~\phi_{2j}\}_{j=1,2,3,...}$ are eigen pairs of equation \ref{eq::1dEigenFunGaussianMeasure} along K direction,  and both $\{\phi_{1i} \}_{i=1,2,3,...}$ and $\{\phi_{2j} \}_{j=1,2,3,...}$ are orthonormal. Then $\{\lambda_{1i} \lambda_{2j}, \phi_{1i} \phi_{2j} \}_{i=1,2,3,...; j=1,2,3,...}$ are eigen paris of equation \ref{eq::twodEigenFun} and $\{\phi_{1i} \phi_{2j} \}_{i=1,2,3,...; j=1,2,3,...}$ are orthonormal. \\
The proof is given in Appendix A.

Then we arrange the obtained eigenvalues $\lambda_{ij} =\lambda_{1i} \lambda_{2j}, ~i=1,2,3... ~and ~ j=1,2,3...$ in a decreasing order, and denote the ordered eigenvalues as $\lambda_k$, and the corresponding eigenfunctions as $\phi_k$, $k=1,2,3...$. These $\lambda_k$'s and $\phi_k$'s will be the solutions of (\ref{eq::twodEigenFun}). To this end, we can approximate the random field Y by keeping first $N_{KL}$ truncation terms of its K-L expansion:
\begin{equation} \label{eq::2deigendecomposition}
    Y(x) = \sum_{k=1}^{N_{KL}} \theta_k \sqrt{\lambda_k} \phi_k(x),
\end{equation}

where $N_{KL}$ is chosen by the desired energy ratio (\ref{eq::energyRatio}).

\begin{figure}[!h]
    \begin{center}
        \begin{tabular}[t]{cc}
          \includegraphics[height=8cm, width = 8cm]{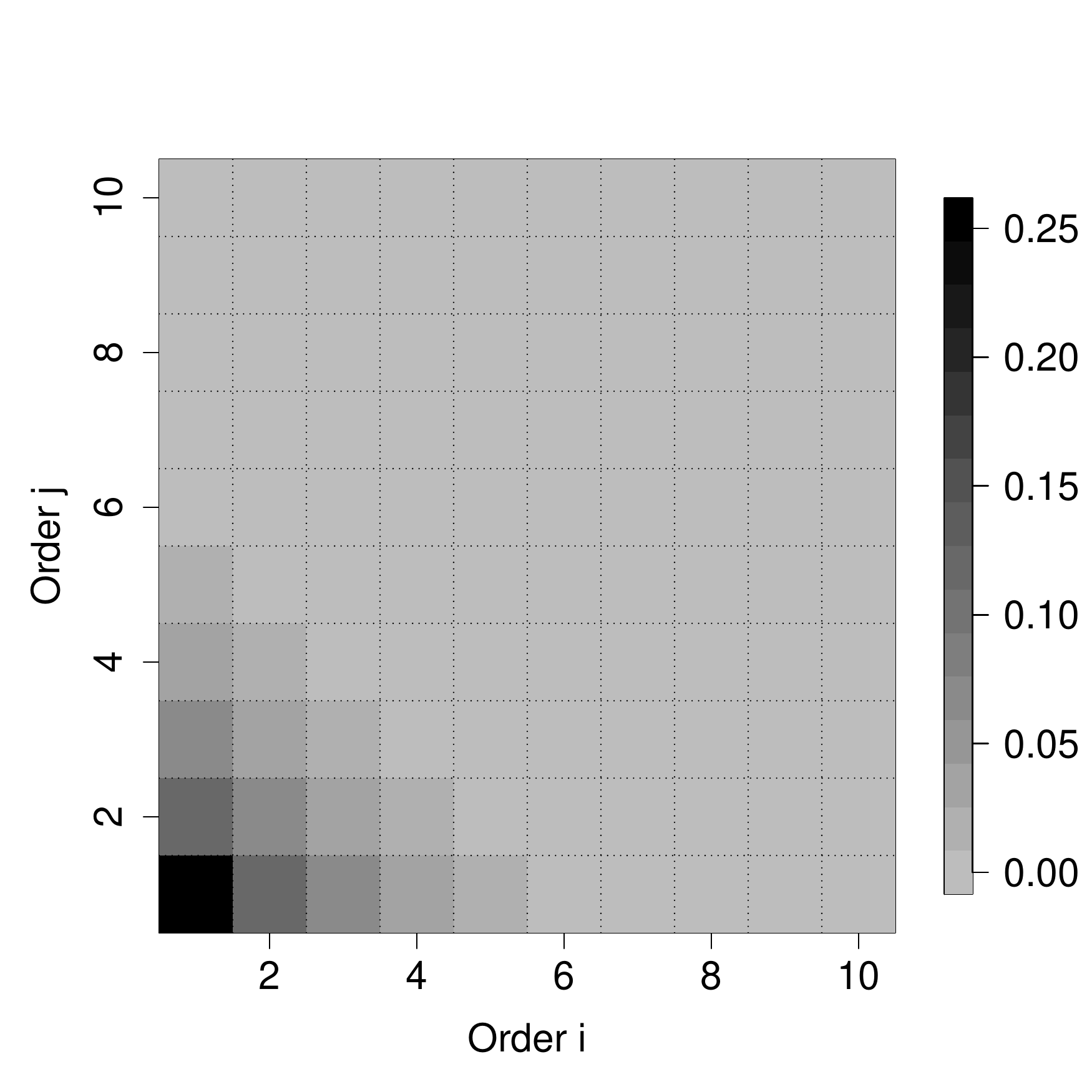}   &  \includegraphics[height=8cm, width = 8cm]{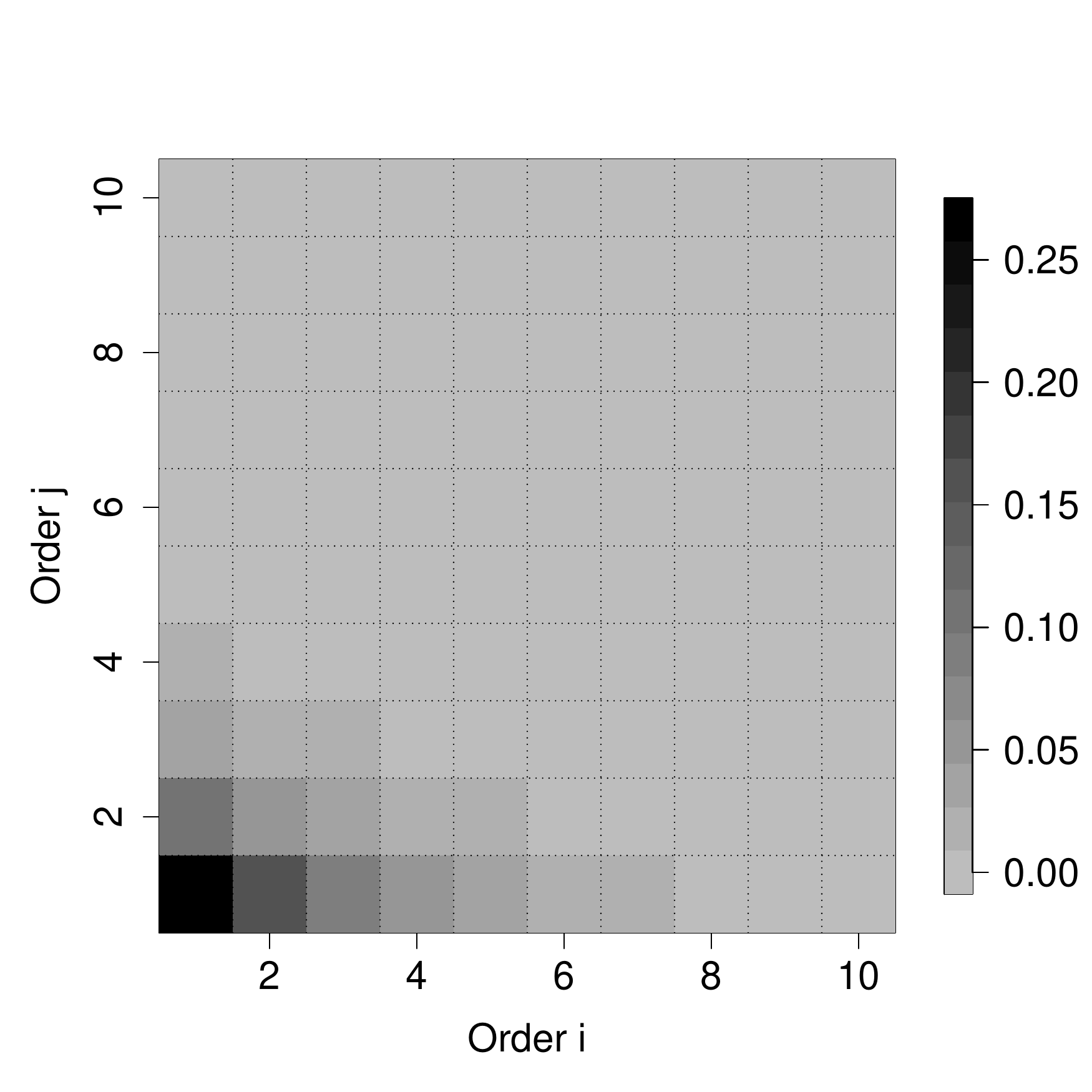}\\
           (a)  &  (b)
        \end{tabular}
    \end{center}
    \caption{(a) Heat map of two dimensional eigenvalues with $l_1 = 0.5$ and $l_2 = 0.5$ ; (b) Heat map of two dimensional eigenvalues with $l_1 = 0.5$ and $l_2 = 0.25$. }
    \label{fig::2deigenvalueSample}
\end{figure}

\begin{figure}[!h]
    \begin{center}
        \begin{tabular}[t]{cc}
          \includegraphics[height=8cm, width = 8cm]{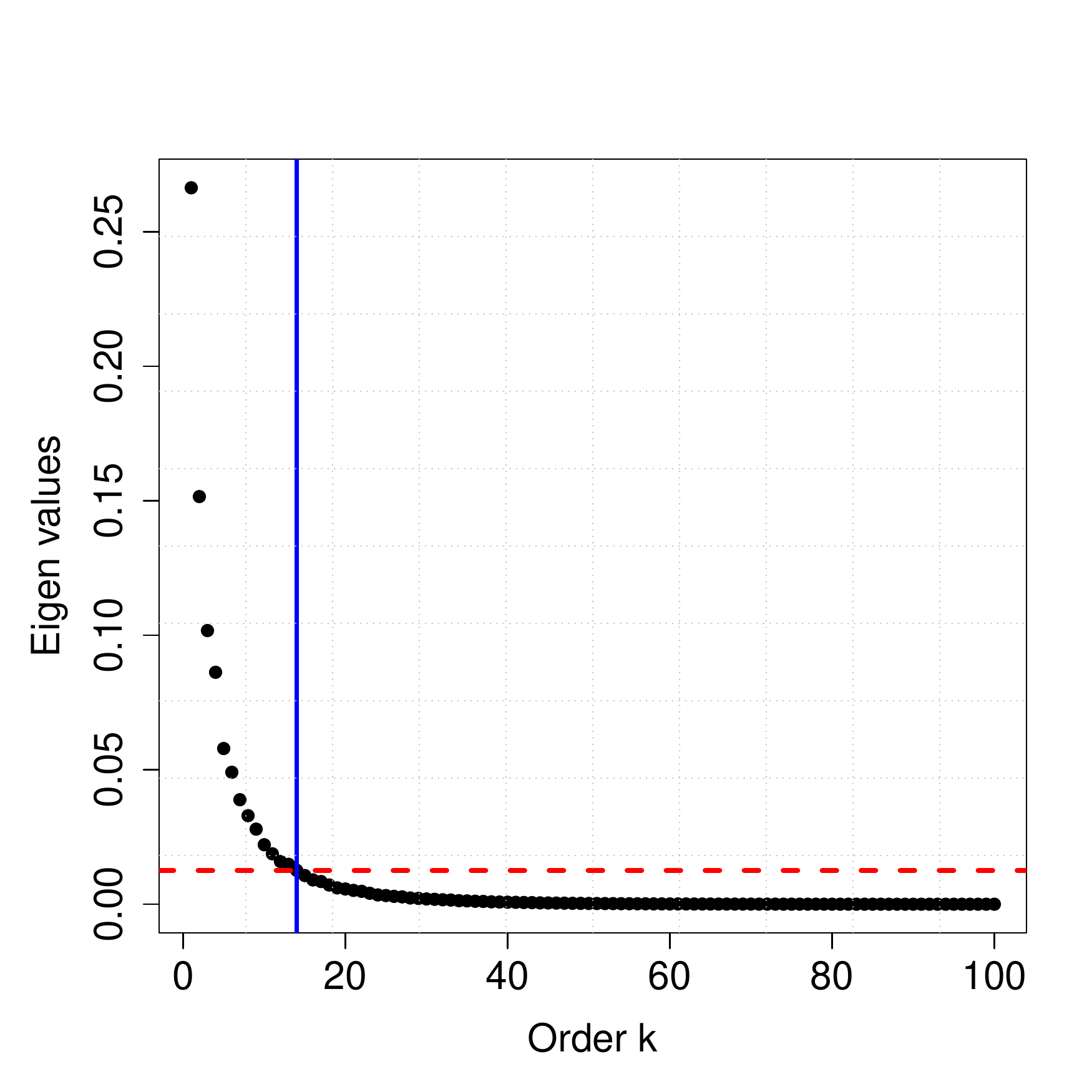}   &  \includegraphics[height=8cm, width = 8cm]{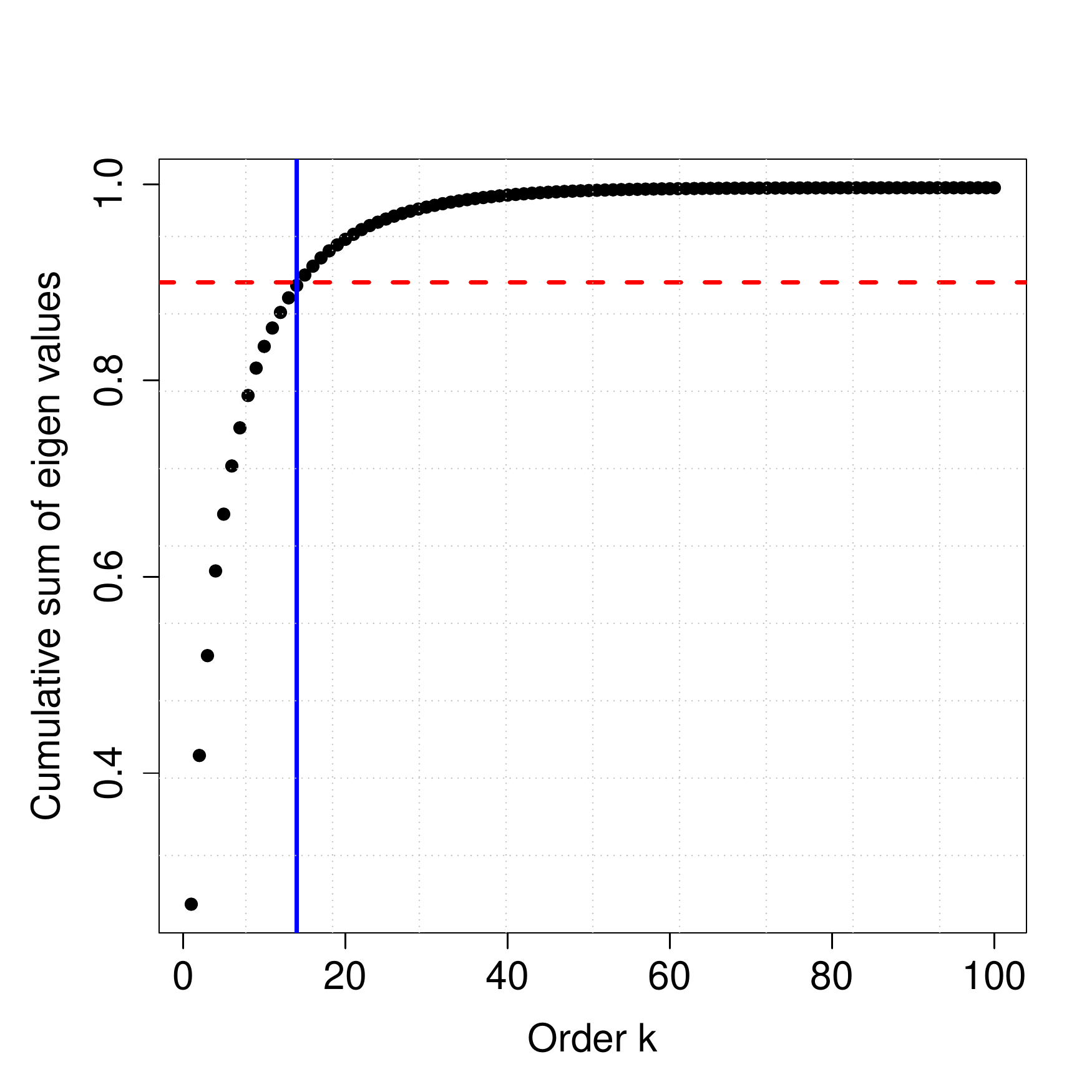}\\
           (a)  &  (b)
        \end{tabular}
    \end{center}
    \caption{(a) Plot of the decreasing ordered 2 dimensional eigenvalues ; (b) plot the cumulative sum of 2 dimensional eigen values. Red dash line marks the 90\% variance cut-off, and blue line indicate 14 terms to capture the 90\% variance. Here we used $l_1 = 0.7$ and $l_2 = 0.4$ in the squared exponential correlation function}
    \label{fig::2deigenvalueordering}
\end{figure}

\begin{figure}[!h]
    \begin{center}
         \begin{tabular}[t]{cc}
          \includegraphics[height=8cm, width = 8cm]{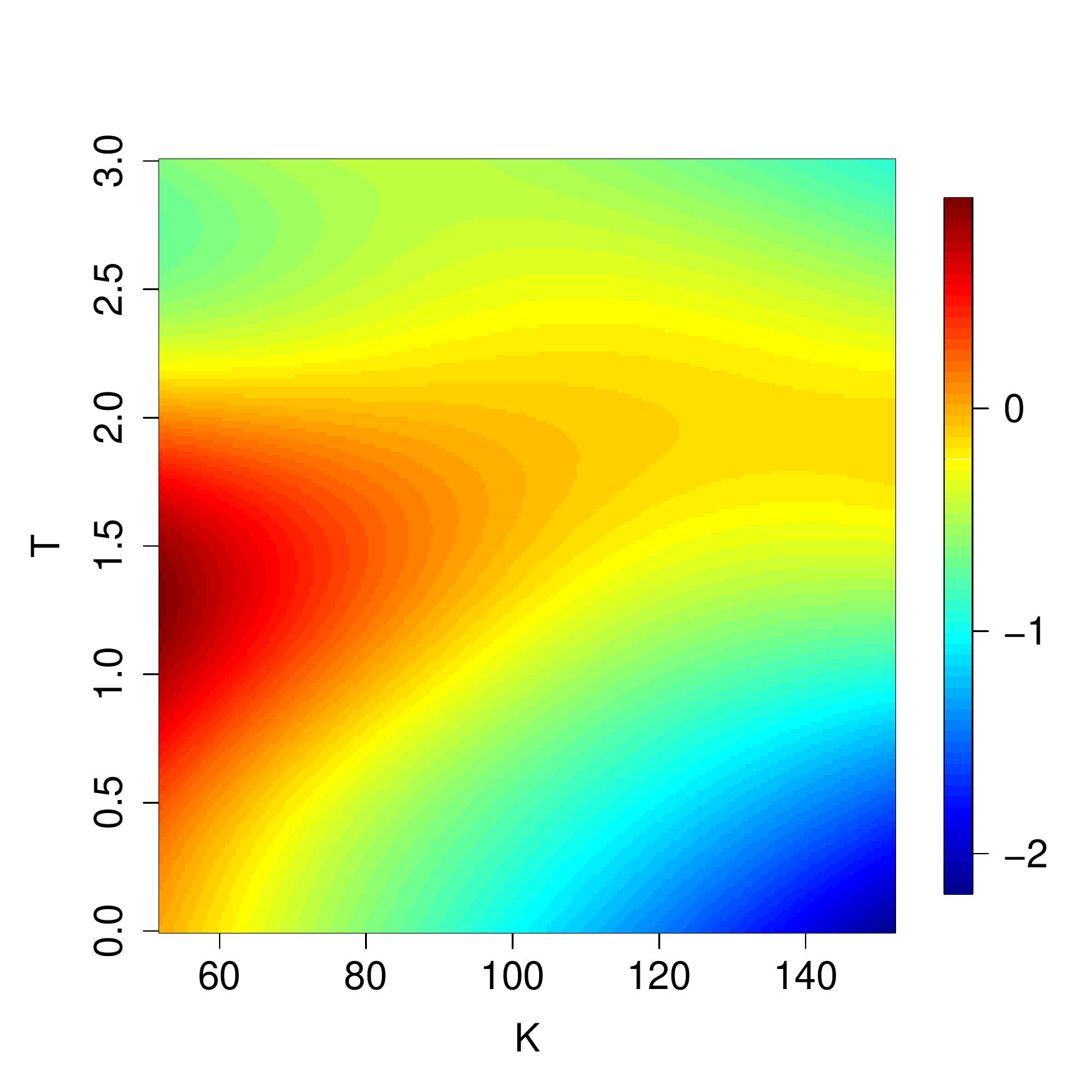}   &  \includegraphics[height=8cm, width = 8cm]{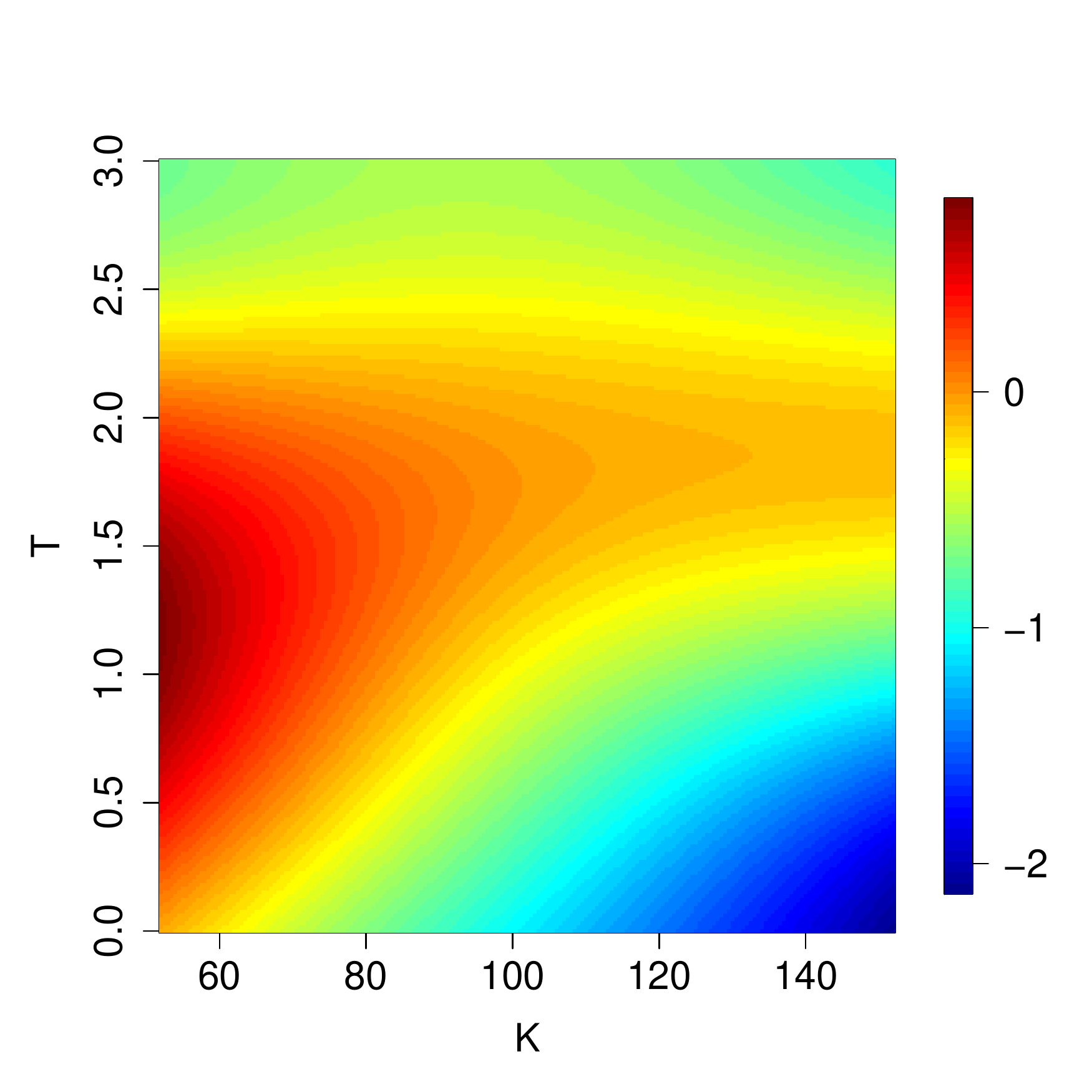}\\
           (a)  &  (b)
        \end{tabular}
    \end{center}
    \caption{(a) Image plot of an arbitrary random field generated by a GP with $l_1 = 0.7$ and $l_2 = 0.4$ ; (b) Image plot of the same random field truncated with 14 terms capturing 90\% of the process variance. }
    \label{fig::2dklsamples}
\end{figure}

To demonstrate the order of $N_{KL}$ terms to be retained in the K-L expansion, we did some numerical experiments to choose a reasonable cut-off point for it. Generally, $N_{KL}$ is selected such that that the energy ratio defined in (\ref{eq::energyRatio}) is at least 90\%. 
Figure (\ref{fig::2deigenvalueSample}) (a) and (b) show the heat map of eigenvalues for two-dimensional GPs with squared exponential kernel. From these plots, we can observe that a small number of eigenvalues on the bottom left corner captures most of the variability of the processes. In Figure (\ref{fig::2deigenvalueordering}) (a) and (b) we plot the decreasing order eigenvalues and the corresponding cumulative values. From these plots, we observe that only a small portion of $N_{KL}$ terms are needed for the truncated K-L process to retain the 90\% of the variability. Figure (\ref{fig::2dklsamples}) shows a comparison of the true surface with the approximated surface obtained from the truncated K-L expansion keeping the first $14$ terms. The approximation appears to be reasonable.

We now have a parametric representation of the random field $Y$ with parameters $l_1, l_2, \sigma, \theta_1, \theta_2, \dots, \theta_{N_{KL}}$, and $Y$ can be approximated analytically if those parameters are known. This saves a significant amount of computational cost when compared to solving the second Fredholm equation \ref{eq::twodEigenFun} numerically at each iteration of the MCMC sampling.

Rewriting the equation (\ref{eq::generalBayesformula}) in terms of expression (\ref{eq::2deigendecomposition}), we have: 

\begin{equation} \label{eq::parametricBayesform01}
     p(\theta, l_T, l_K, \sigma_Y, \sigma_{\epsilon} | \V^{obs}) \propto p(\V^{obs}|\theta, l_T, l_K, \sigma_Y, \sigma_{\epsilon})p(\theta, l_T, l_K, \sigma_Y^2, \sigma_{\epsilon})
\end{equation}

\subsection{The likelihood and Prior Distributions}

The parametric likelihood can be obtained from (\ref{eq::noiseModel}) and (\ref{eq::2deigendecomposition}) as $V_{ij}^{obs} = V(T_i, K_{ij}, \theta, l_1, l_2, \sigma_Y) + \epsilon_{ij}$, where the model option value $V$ is now a realization from a forward model with parameters $\theta, l_1, l_2, \sigma_Y$ as its input variables. Inverse Gamma distribution is used for the prior distribution of  $\sigma_{\epsilon}$, i.e., $\sigma_{\epsilon} \sim \text{IG}(a_{\epsilon},b_{\epsilon})$. Integrating out $\sigma_{\epsilon}^2 $, yields the marginal likelihood as 
\begin{equation}
\label{eq::like}
p(\V^{obs}|\theta, l_T, l_K, \sigma_{Y}^2) \propto \frac{\Gamma(a_{\epsilon}+n/2)}{[b_{\epsilon}+\frac{1}{2}|| \V^{obs} - V^{}(\theta, l_T, l_K,\sigma_{Y}^2)||^2]^{(a_{\epsilon}+n/2)}}
\end{equation}

For K-L expansion of a Gaussian process, $\theta$ follows a Multivariate Normal Distribution $\text{MVN}(0, I_{N_{KL}})$. 
The degree of smoothness also depends on $ l_1, l_2 \text{ and } \sigma_Y$. Without any prior information about the parameter, we use uniform priors for $l_1$ and $l_2$ with carefully chosen upper and lower bounds. The bounds are chosen such that the degree of smoothness is within the acceptable range. Lastly, the prior distribution for $\sigma_Y$ is taken to be Gamma$(a_s, b_s)$. The hyper-parameters $a_s$ and $b_s$ can be chosen based on the implied volatility data.  
From \ref{eq::parametricBayesform01} and \ref{eq::like}, the posterior distribution is given by
\begin{align}
     \label{eq::parametricBayesform_Final}
    \pi(\theta, l_T, l_K, \sigma_Y, \sigma_{\epsilon} )= p(\theta, l_T, l_K, \sigma_Y, \sigma_{\epsilon} | \V^{obs}) \propto & \frac{\Gamma(a_{\epsilon}+n/2)}{[b_{\epsilon}+\frac{1}{2}|| \V^{obs} - V^{}(\theta, l_T, l_K,\sigma_{Y}^2)||^2]^{(a_{\epsilon}+n/2)}}     \\ 
      &\times exp(-\frac{1}{2}\theta'\theta) \times (\sigma_Y^2)^{a_s-1)}exp(\sigma^2/b_s)
\end{align}

\section{Bayesian Computation Using MCMC}\label{sec::computation} 

As the likelihood \eqref{eq::like} contains the forward simulator $V$ which is the solution to the PDE (\ref{eq::DupireE}), the posterior (\ref{eq::parametricBayesform01}) is not tractable. So, we use an MCMC based method to sample from the posterior and the Bayesian inference about the model parameters is based on these samples. At each step of the MCMC method, the forward model needs to be solved numerically to evaluate the likelihood. In this section we first discuss the numerical solver for the forward model then we discuss the MCMC sampling algorithm.

\subsection{Numerical Solver for the Option Pricing Model}
\begin{figure}[!h] 
    \begin{center}
        \begin{tabular}[t]{cc}
          \includegraphics[height=8cm, width = 8cm]{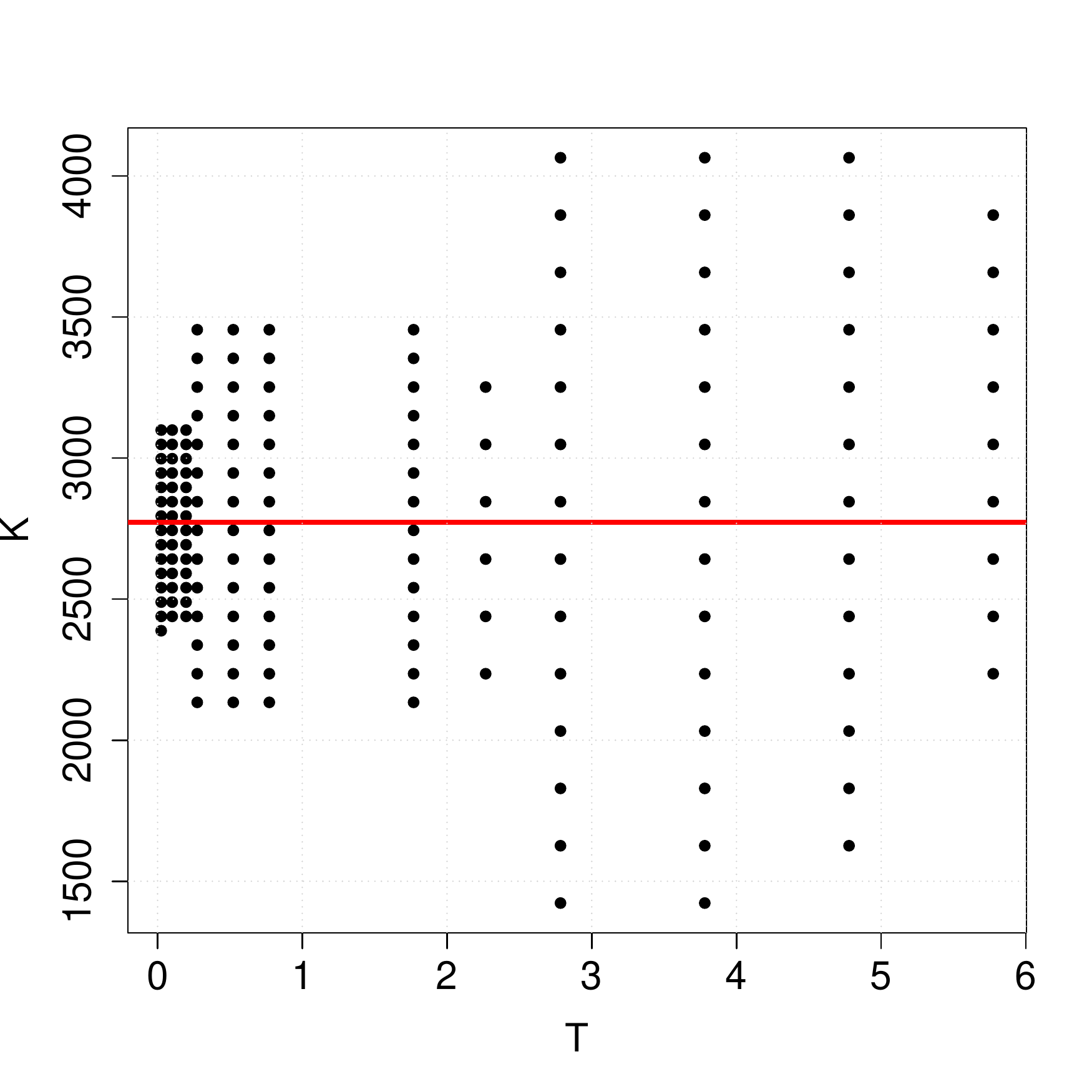}   &  \includegraphics[height=8cm, width = 8cm]{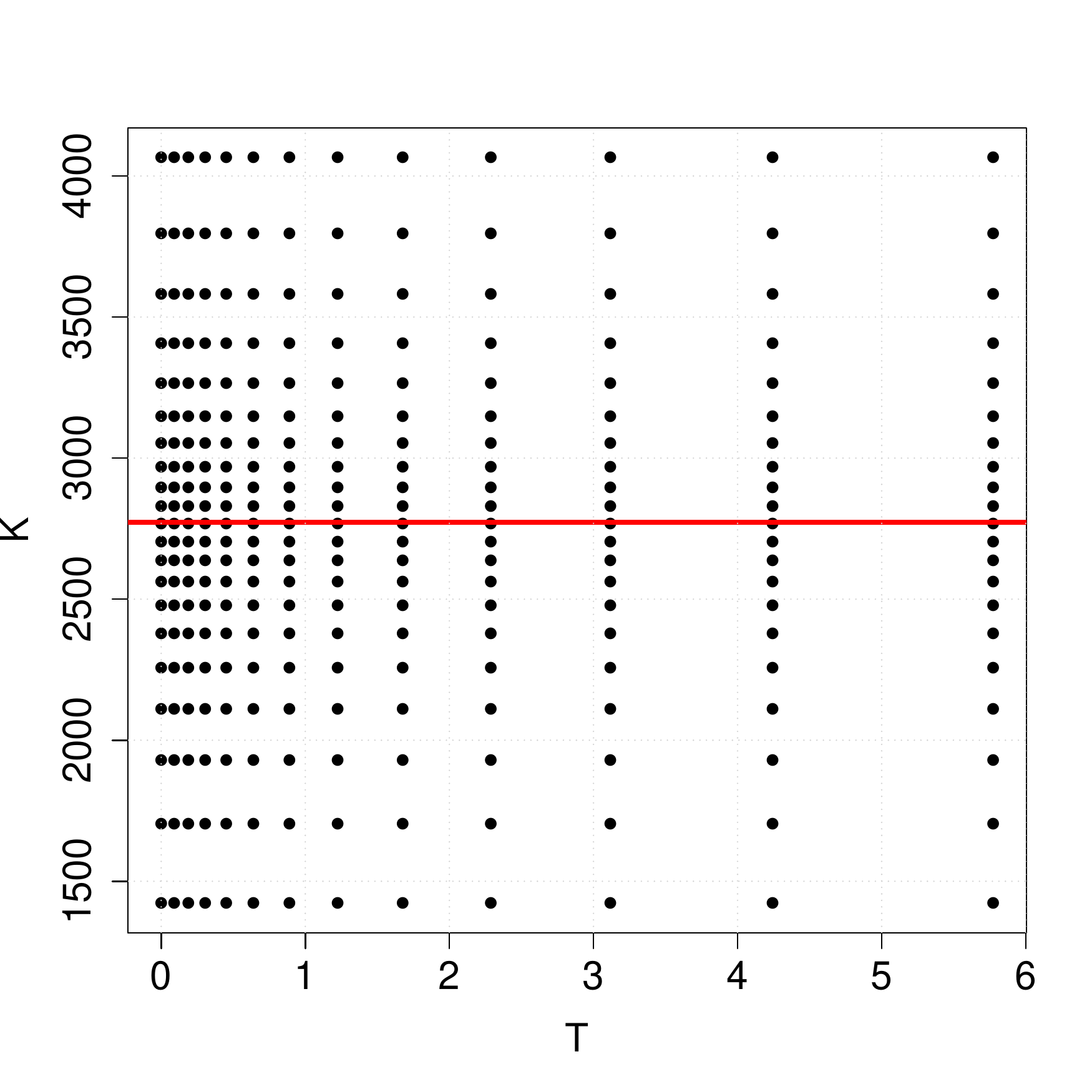}\\
           (a)  &  (b)
        \end{tabular}
    \end{center}
    \caption{(a) Quoted strikes and maturities of option data in section \ref{sec::numericalRealData} Case 5; (b) Non-uniform mesh for the finite element method.}
    \label{fig::meshGrid}
\end{figure}

There is no closed-form solution to \ref{eq::DupireE}, hence numerical methods are used to solve it. The traditional numerical solver uses a finite-difference method in a grid to solve the forward model and uses a post interpolation method to predict the model option prices corresponding to the market data. We solve the Dupire's PDE \ref{eq::DupireE} using a finite element method (FEM) in the K direction following \citep{hirsa2012computational}. There are many advantages to using the FEM in our context.
In the FEM the solution can be obtained at any grid as it is based upon a piecewise representation of the solution in terms of specified basis functions. So no post interpolation is necessary for the finite element method. The FEM is also flexible in choosing the mesh grid for the basis functions according to the distribution of the sampling points. In particular, options are heavily traded near at-the-money with shorter maturities, meanwhile, the option prices in that area are more sensitive to volatility. Hence, it is natural to use a non-uniform mesh grid \citep{hirsa2012computational} along the K direction to achieve high accuracy in solving \ref{eq::DupireE} with a small number of mesh points. Figure \ref{fig::meshGrid} shows the mesh scheme we implemented in the case studies in section \ref{section::caseStudies}.
The FEM also provides stabler coarse grid approximations in the first stage of TSAM (see details in the next subsection). In the finite element method we first derive the weak form for (\ref{eq::DupireE}) and then express the true solution as a piece-wise linear combination of quadratic basis function defined on non-uniform mesh along K direction. The resulting linear system of ODEs is then solved using a finite difference method \citep{crank_nicolson_1947} in a regular time grid. The details the derivation of the system of equation for the numerical solver can be found in Appendix B.

\subsection{The MCMC Sampling Algorithm}

In an MCMC-based sampling for posterior probability distribution \ref{eq::parametricBayesform01}, the dimension of the posterior is still high even after dimension reduction using K-L expansions. Also, there is a high computational cost in repetitive solving of the forward model numerically on a fine grid. Here, we use a two-stage adaptive Metropolis algorithm (TSAM) \citep{Mondal2021ATS} to sample from the posterior distribution, which combines two quite powerful ideas in the MCMC literature $-$ adaptive Metropolis sampler and two-stage Metropolis-Hastings sampler. Let us denote $\xb =(\theta, l_1, l_2, \sigma_Y)$ to be all the parameters of the posterior distribution. In the first stage of the algorithm, an adaptive proposal is used based on the previously sampled states, and the corresponding acceptance probability is computed based on an approximated inexpensive target density $\pi^{*}$, given by
\begin{align}
     \label{eq::parametricBayesform_algorithm}
    \pi^*(\xb)= p(\theta, l_T, l_K, \sigma_Y, \sigma_{\epsilon} | \V^{obs}) \propto & \frac{\Gamma(a_{\epsilon}+n/2)}{[b_{\epsilon}+\frac{1}{2}|| \V^{obs} - V^{*}(\theta, l_T, l_K,\sigma_{Y}^2)||^2]^{(a_{\epsilon}+n/2)}}     \\ 
      &\times exp(-\frac{1}{2}\theta'\theta) \times (\sigma_Y^2)^{a_s-1)}exp(\sigma^2/b_s),
\end{align}
where $V^*$ is the approximated model option price computed by the finite element method using a coarser grid. The expensive posterior distribution $\pi(\xb)$, defined in equation \ref{eq::parametricBayesform_Final}, with $V$  obtained from a finer grid, is evaluated in the second stage only if the proposal is accepted in the first stage. The adaptive nature of the algorithm guarantees faster convergence of the chain and very good mixing properties. On the other hand, the two-stage approach helps in rejecting the bad proposals in the inexpensive first stage, making the algorithm computationally efficient.  We describe the TSAM algorithm as below:

\begin{algorithm}[H] \label{algo::TSAM}
\caption{Two Stage Adaptive Metropolis (TSAM)}
    \begin{enumerate}
            \item A candidate $\xb^*$ is sampled from the proposal distribution  $q_t(.|\xb_0, \xb_1, ..., \xb_{t-1})$, which is a Gaussian distribution with mean $\xb_{t-1}$ and covariance $C_t$ defined as
            \beq
            \label{ct}
            C_t = 
            \begin{cases}
            C_0 & \text{if} \ t <t_0, \\
            s_d {\rm cov}(\xb_0, \xb_1, ..., \xb_{t-1}) + s_d \epsilon \bs{I}_d & \text{if} \ t\geq t_0.
            \end{cases}
            \eeq
            Here $s_d$ is a parameter that depends only on the dimension $d$, $\bs{I}_d$ denotes the $d$-dimensional identity matrix, and $\epsilon$ is a small constant which ensures that $C_t$ does not become singular. In order to start, an arbitrary, strictly positive definite initial covariance $C_0$ is selected according to the best prior knowledge.  $t_0 \ (>0)$ is the length of an initial period for which $C_0$ is used for the proposal distribution.
            
            \item  In the first stage the candidate $\xb^*$ is screened and passed to the second stage with acceptance probability given by
            \beq
            \label{stage1ap}
            \alpha_1(\xb_{t-1}, \xb^*) = \min\Big(1, \frac{\pi^*(\xb^*)}{\pi^*(\xb_{t-1})}\Big).
            \eeq
            This is equivalent of taking the final proposal as 
            \beq
            \xb = 
            \begin{cases}
            \xb^* & \text{with probability} \ \alpha_1(\xb_{t-1}, \xb^*), \\
            \xb_{t-1} & 
            \text{with probability} \ 1- \alpha_1(\xb_{t-1}, \xb^*).
            \end{cases}
            \eeq
            \item Accept $\xb$ as the $t$-th sample from $\pi$ with the second stage acceptance probability 
            \beq
            \label{stage2ap}
            \alpha_2(\xb_{t-1}, \xb) = \min \Big(1, \frac{\pi(\xb)\pi^*(\xb_{t-1})}{\pi(\xb_{t-1})\pi^*(\xb)} \Big).
            \eeq
    \end{enumerate}
\end{algorithm} 
Note that the above sampling algorithm produces a chain that does not have the Markov property, but in \cite{Mondal2021ATS} we showed the chain maintains the ergodicity property.

\section{Applications to Simulated and Real Data}\label{section::caseStudies}
\begin{figure}[!h]
\begin{center}
\includegraphics[scale=0.60]{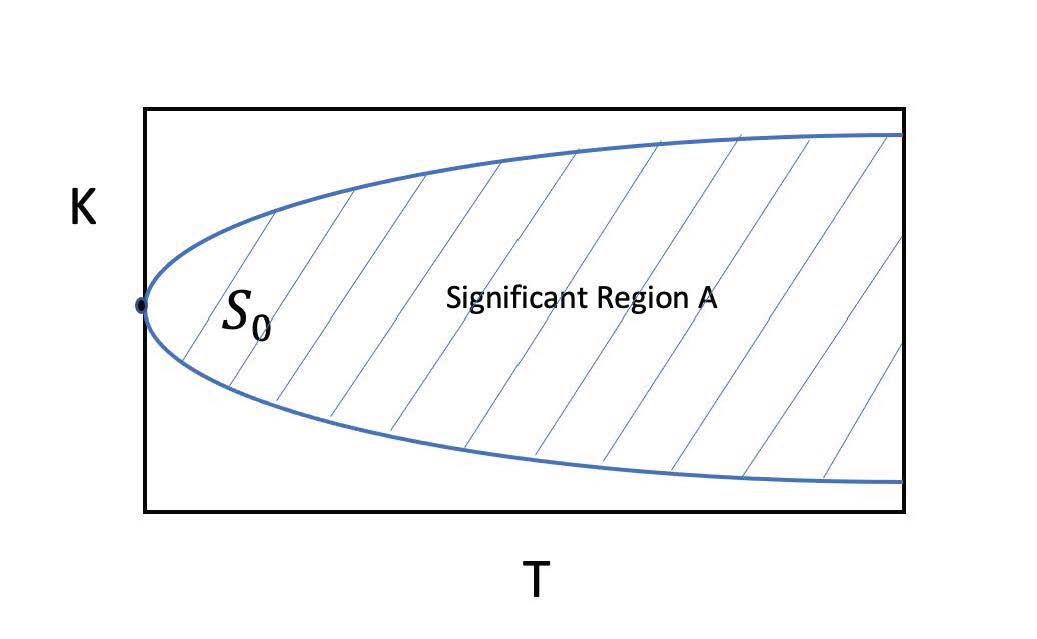}
\end{center}
\caption{The option premiums in shaded area are more sensitive to local volatility. }
\label{fig::lvSignificant}
\end{figure}

We apply our methodology to three synthetic examples and two market data examples. In the simulation studies, without loss of generality, we rescale the domain of strike price and maturity time to a unit square $[-0.5,0.5] \times [-0.5, 0.5]$ to generate local volatility samples. Finding a proper Gaussian measure variance $\sigma_{\mu}$ is critical when implementing an analytical K-L dimension reduction methodology. From empirical studies, we found that $\sigma_{\mu}=0.68$ balances the rich representation of the eigenfunctions and the total percentage of variance on the unit square domain. We divide the data sets into training and validation groups for both synthetic and market data. For synthetic data, we validate our method by directly comparing it with the known local volatility surfaces. However, for the real market data example, the true volatility surface is unknown, thus we validate our methodology through a comparison of repriced options with the validation data set. The posterior samples for local volatility surfaces are used to study the uncertainties of the corresponding option prices and to forecast options with longer maturities.
According to \cite{Coleman1999ReconstructingTU}, the sensitivity of option prices to volatility is not uniform over the region $[K_{min}, K_{max}]\times[0,T_{max}]$. Across all maturities, options with strike prices that are near the money are most sensitive to volatility, while options that are further in the money or out of the money will be less sensitive to volatility. For short maturities, the intrinsic values of the option will dominate the option price with little time value effect. For example, if K is far above $S_0$, that has little chance to move into the money hence the option value is likely to expire worthlessly; if K is far below $S_0$, then it has little chance to move out of the money hence the option value is close to $S_0-K$. For those two scenarios, volatility is not likely to affect option prices. On the other hand, for longer maturities, the time value starts to dominate the option values and the option values with K far from $S_0$ become more and more sensitive to volatility compared to short maturities because of the diffusion assumption in \ref{eq::underlyingProcess}. For example, if T is very long, then the stock price can have more chance to move in and out of money and the frequency depends on the value of volatility. Figure \ref{fig::lvSignificant} provides a general idea of the significant region A, where the option price is more sensitive to volatility. Thus, in the best scenario, we can reconstruct the local volatility well in region A from the market option data. This phenomenon is indeed in agreement with the options market. From figure \ref{fig::meshGrid}, we can see that the traded options form a similar shape as region A. Thus, we do not need to match all of the market prices of options that are far from the money since they may increase the computational complexity of the model considerably while only having a small effect on the prices and Greeks of exotic options that are calculated using the local volatility. The calibrated model is used to predict non-publicly traded long-term options. In our case, we predict the validation data set to demonstrate the predictive performance of our calibrated model. The longer the maturity, the larger the uncertainty level we should expect for the predicted option prices.

\subsection{Numerical Results for Synthetic Data}
We first use synthetic data to illustrate our methodology for estimating the unknown local volatility function from a finite number of market option data. We generate the data with finer grids $N_K=201$ and $N_T=201$ in a uniform mesh and then retain the data on a nonuniform mesh, as mentioned in section \ref{sec::computation}, and consider this as our observed data. While using the TSAM sampler to sample from the posterior we use a coarser grid of $11 \times 11$ for the FEM in the first stage and a finer grid of $51 \times 51$ in the second stage. The Markov chain using the traditional Metropolis-Hastings sampler fails to converge to the target distribution even with $500000$ iterations, for most of the synthetic cases when we retain $18$ or more terms in the K-L expansion. The tuning of the jump size in high dimensional is the biggest challenge here. The TSAM not only converges faster (within $50000$ iterations) to the target distribution due to the adaptive nature of the proposal, the use of the course-grid solver in the first stage to filter the bad proposals makes the computation much faster. In our simulation studies, the TSAM is at-least 10 times faster than the adaptive Metropolis. \\

\begin{figure}[!h] 
    \begin{center}
        \begin{tabular}[t]{cc}
            \includegraphics[height=7.5cm, width = 7.5cm]{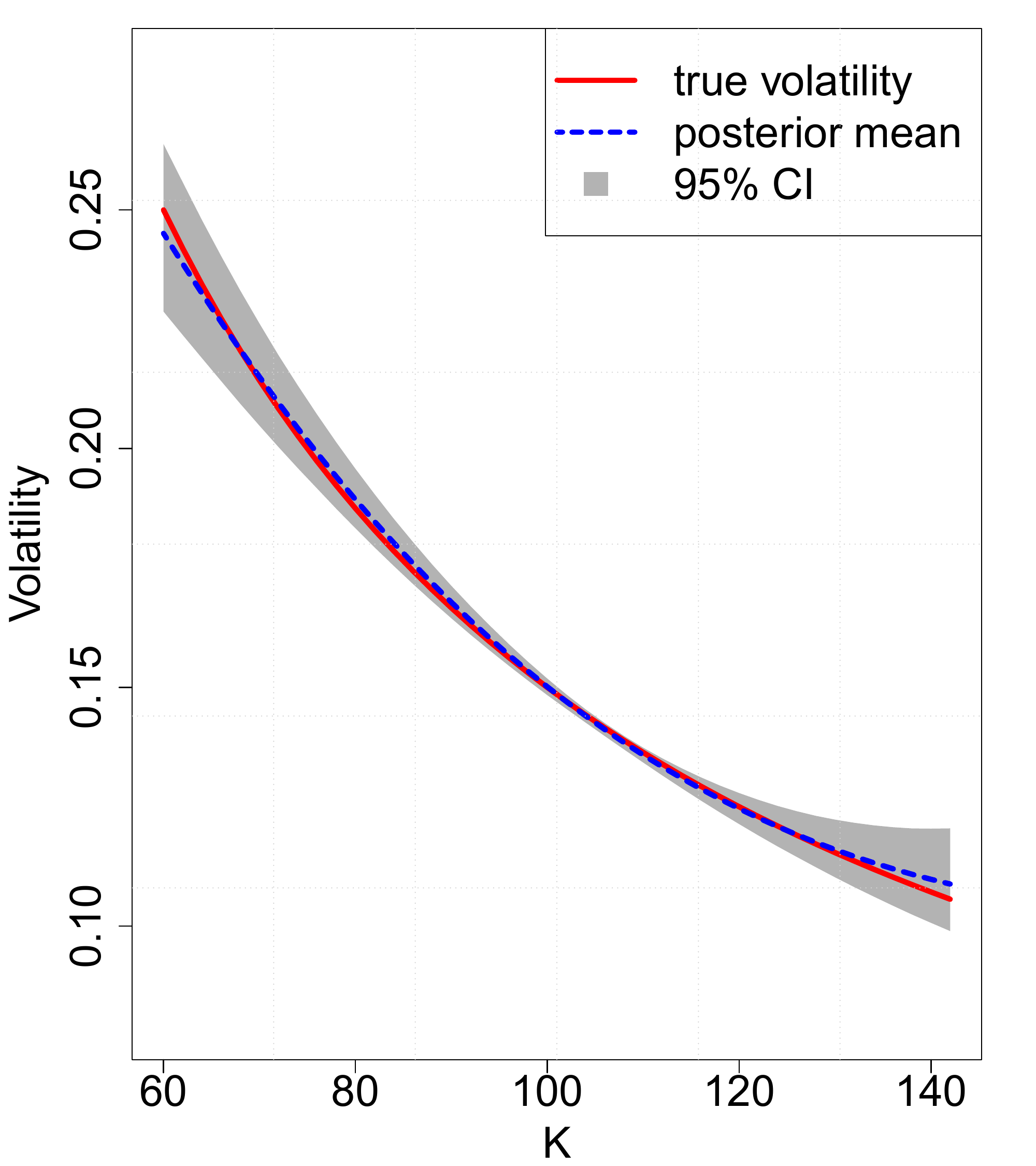}   &   \includegraphics[height=8.5cm, width = 8.5cm]{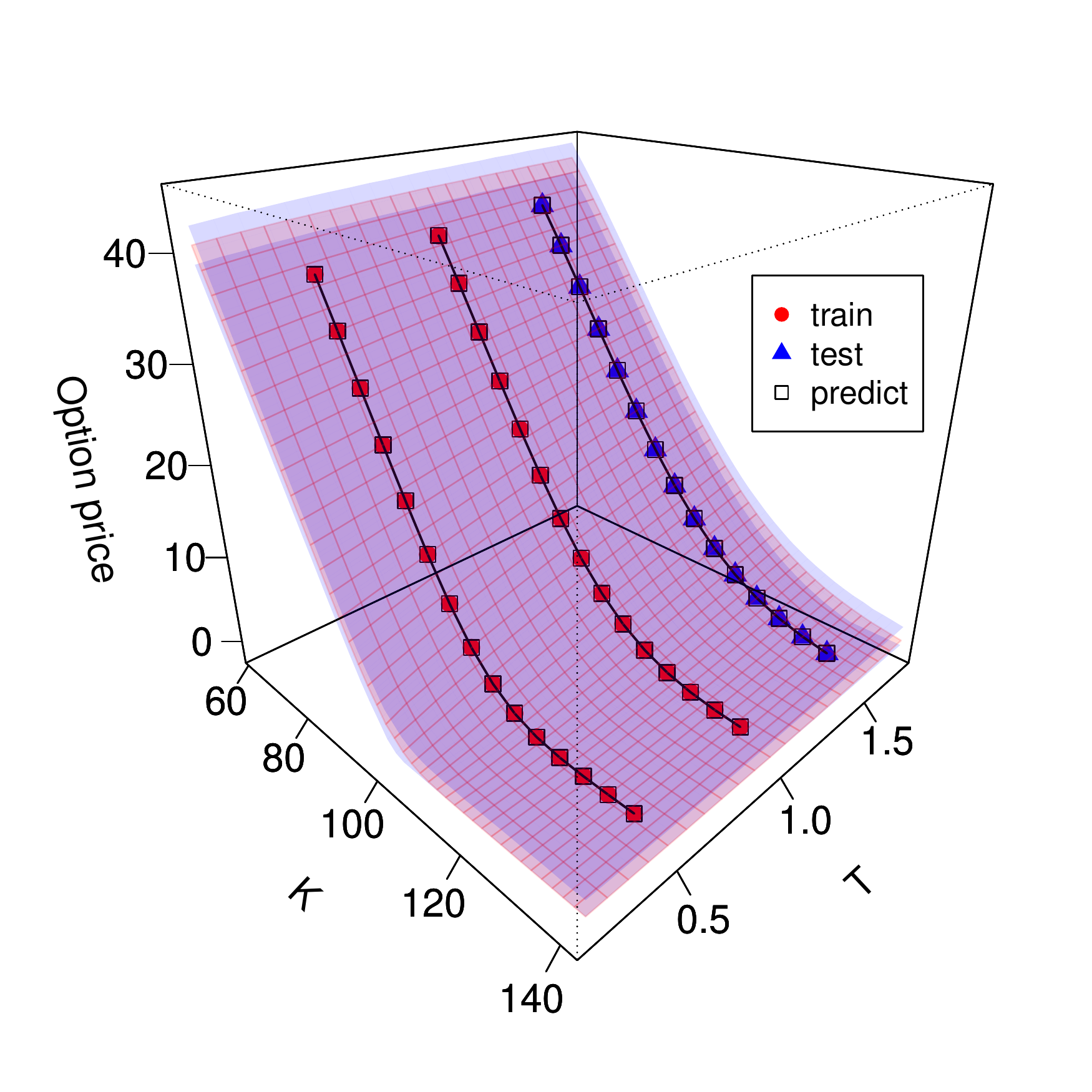}\\
              (a)  &  (b) 
        \end{tabular}
    \end{center}
    \caption{(a) The solid red curve is the benchmark volatility from case 1, the blue dashed curve is the posterior mean and the greyed region represents the 95\% credible band; (b) The red dots are the observed market data, the blue triangles are corresponding to the held-out data, and the unfilled squares are the corresponding posterior predictive mean. The light blue surfaces are the corresponding 95\% posterior prediction bands.}
    \label{fig::1dlv}
\end{figure}

For the first synthetic example (Case 1) we use a constant elasticity of variance model where local volatility depends only on strike prices, as mentioned in \cite{lagnado1997technique}.

\begin{equation}\label{eqn::1dlv}
\sigma (K,T) = K/15.
\end{equation}
We use the same spot price $S_0$ = 100, risk free interest rate = 0.05, and the dividend rate q = 0.02. To be consistent, we use these same parameter values for the next two synthetic cases (Case 2 and 3). The synthetic option data is generated with $T \in [0.5, 1.5]$ and $K \in [60,140]$, and using \ref{eqn::1dlv} as the underlying true local volatility function. In figure \ref{fig::1dlv} (b), the red dots corresponding to $T=0.5$ and $T= 1$ represent the option data used for model calibration and the blue triangles corresponding to $T=1.5$ represent the option data used for validation. The prior for the correlation length, $l$, is taken to be a uniform distribution over $[0.5, 1]$. We keep 14 terms in the truncated K-L expansion, i.e.,  $N_{KL}= 14$. This is sufficient to capture the $90\%$ variability of the local volatility. The prior for $\sigma_Y^2$ is assumed to be a gamma distribution with hyper-parameters $a_s = 3$ and $b_s =2$. The same priors are used for the next two simulation studies, viz., synthetic cases 2 and 3. The TSAM is used to draw 100,000 samples from the posterior. We thin the chain by keeping every 10th sample after a 10,000 burn-in period. From figure \ref{fig::1dlv}(a), we can see that our posterior mean is very close to the true volatility curve. The 95\% credible bands also contain the true volatility curve. In figure \ref{fig::1dlv}(b), the unfilled black squares represent the repriced options for the training data and forecasted options for the validation data. These are computed using the mean of the posterior predictive distribution. We can see that the predictive mean matched both the training and validation data very closely. The corresponding 95\% credible bands for option prices also contain the benchmark case for both the training and validation data. The uncertainties in the option prices are narrower than the ones in local volatility, which is because that the strikes far from at-the-money are less sensitive to option prices than at-the-money strikes.

In the second synthetic example (case 2), we generate the option data using the following local volatility function as used in \cite{hirsa2012computational}, which depends on both K and T:

\begin{equation}\label{eqn::2dlv}
\sigma (K,T) = 0.3*exp(-T)(\frac{100}{K})^{0.2}. 
\end{equation}

\begin{figure}[H] 
    \vspace{-20mm}
    \begin{center}
    \includegraphics[scale=0.65]{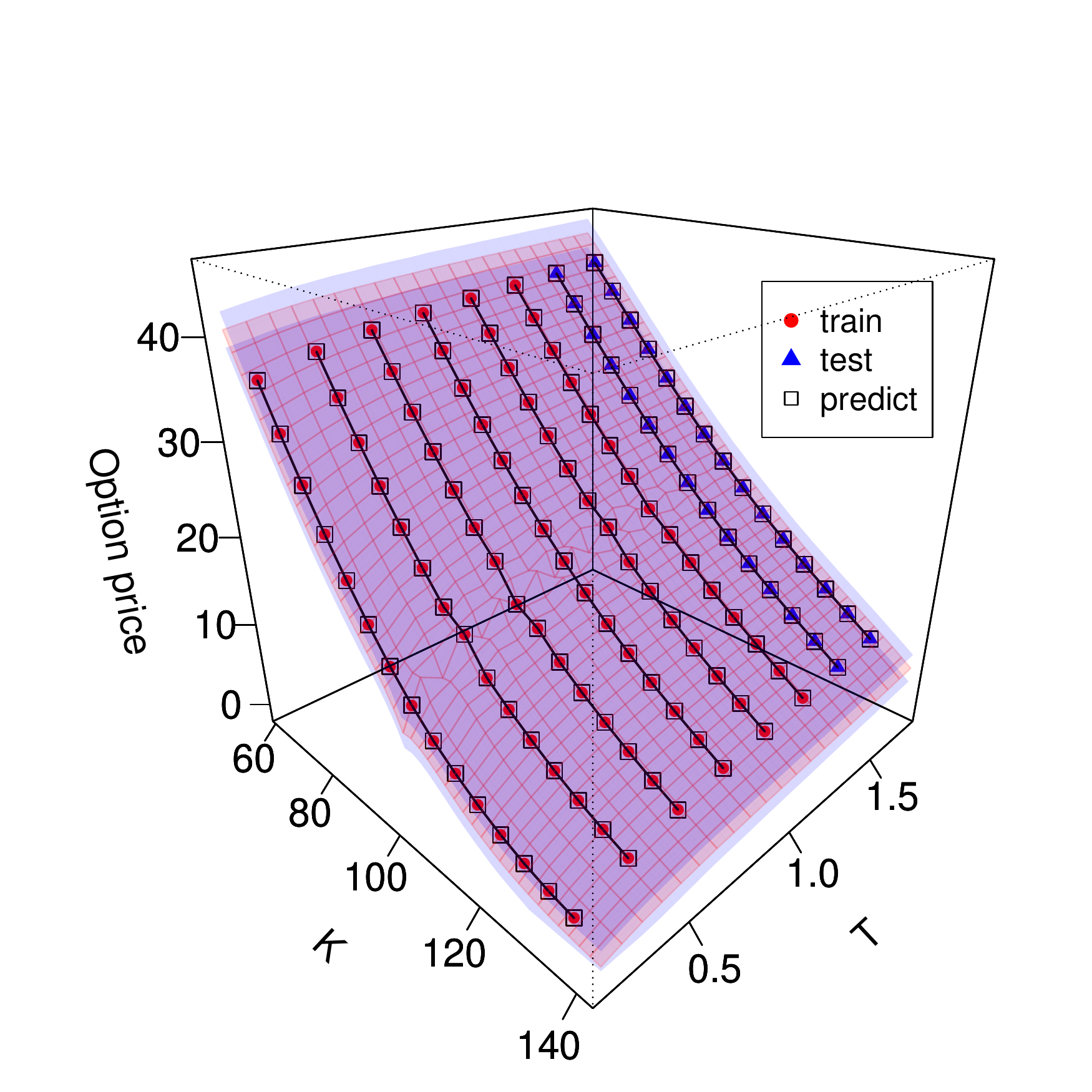}
    \end{center}
    \vspace{-5mm}
    \caption{The red dots are the observed market data from Case 2, the blue triangles are the corresponding held-out data, and the unfilled squares are the corresponding posterior predictive mean. The light blue surfaces are the corresponding 95\% posterior prediction bands.}
    \label{fig::2dlvDataPredcition}
\end{figure}

\begin{figure}[H]
    \begin{center}
        \vspace{-10mm}
        \begin{tabular}[t]{cc}
              \includegraphics[height=8.5cm, width = 8.5cm]{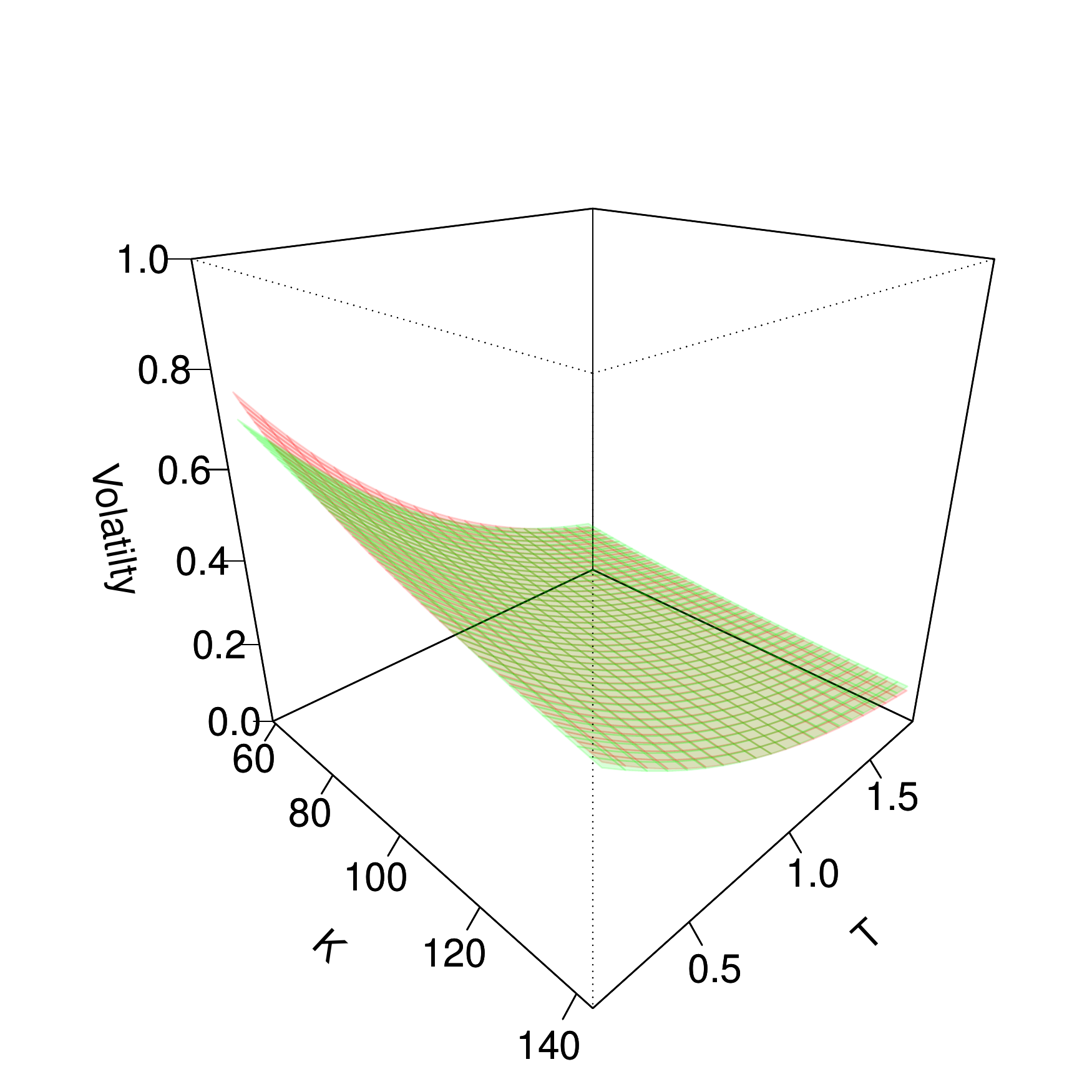}   &  \includegraphics[height=8.5cm, width = 8.5cm]{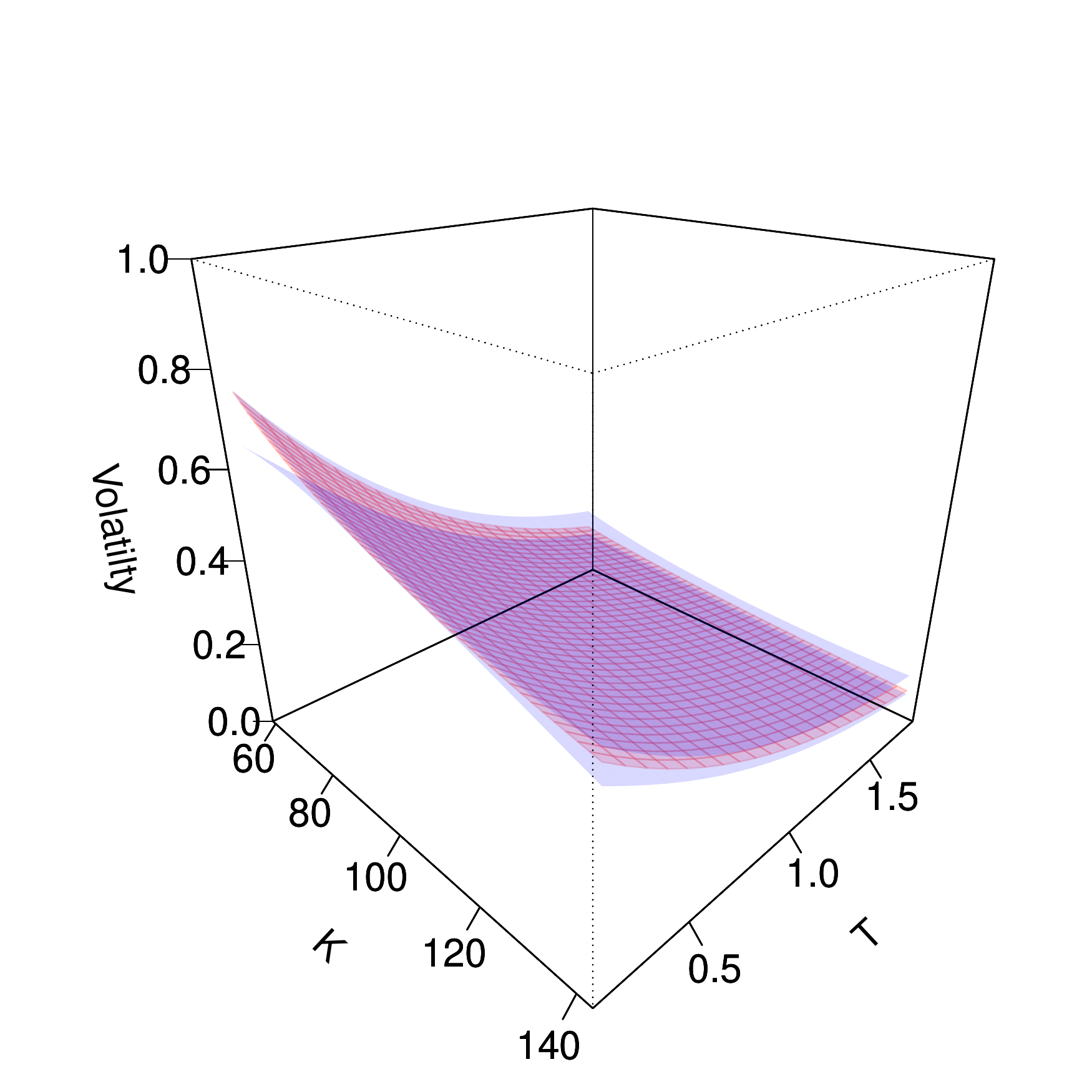}\\
               (a)  &  (b)
        \end{tabular}
    \end{center}
    \vspace{-5mm}
    \caption{(a) The red surface is the true volatility surface from Case 2 and the green surface is the posterior mean; (b) The red surface is the true volatility surface and the blue surfaces represent the 95\% credible surface bands.}
    \label{fig::2dlv}
\end{figure}

The synthetic option data is generated with $T \in [0.25,2]$ and $K \in [65,135]$, and using \ref{eqn::2dlv} as the underlying true local volatility surface. We generate a total of 120 option prices and use options with the first six maturities for model calibration and the rest options for validation, they are shown as red circles and blue triangles respectively in figure \ref{fig::2dlvDataPredcition}.
The prior for both $l_1$ and $l_2$ are taken to be uniform distribution over $[0.5,1]$. We use the truncated K-L expansion in two dimensions as described in section \ref{subsubsec::analytical K-L}, and we keep $22$ K-L terms for retaining $90\%$ of variability. The TSAM is used to draw 300,000 samples from the posterior. We thinned the chain by keeping every $10$th sample after a $50,000$ burn-in period. From figure \ref{fig::2dlvDataPredcition}, we can see that our posterior predictive mean for the option prices is very close to the validation data, and the validation data is contained inside the $95\%$ credible bands. Figure \ref{fig::2dlv} (a) shows the comparison of the benchmark volatility surface and the corresponding posterior mean. As we can see the posterior mean is very close to the true volatility surface. Figure \ref{fig::2dlv} (b) shows the benchmark volatility surface is contained by the $95\%$ credible bands. The uncertainties in local volatility surface are not reflected in repriced European options as the sensitivity of option prices to the local volatility is very low outside the significant region A. For example, in short maturity and strike far away from spot price the volatility has high uncertainty but the transferred uncertainty in the option price is very negligible. The reason being, regardless of what the volatility is, the option will expire worthless almost surely. However, this is only true for European options which are not as sensitive as other exotic options that are more heavily dependent on volatility.
Note that in Case 1 and Case 2 the volatility surface is generated using synthetic parametric models but, in both cases, the Gaussian process can accurately model the volatility surfaces.

\begin{figure}[H] 
    \vspace{-20mm}
    \begin{center}
    \includegraphics[scale=0.65]{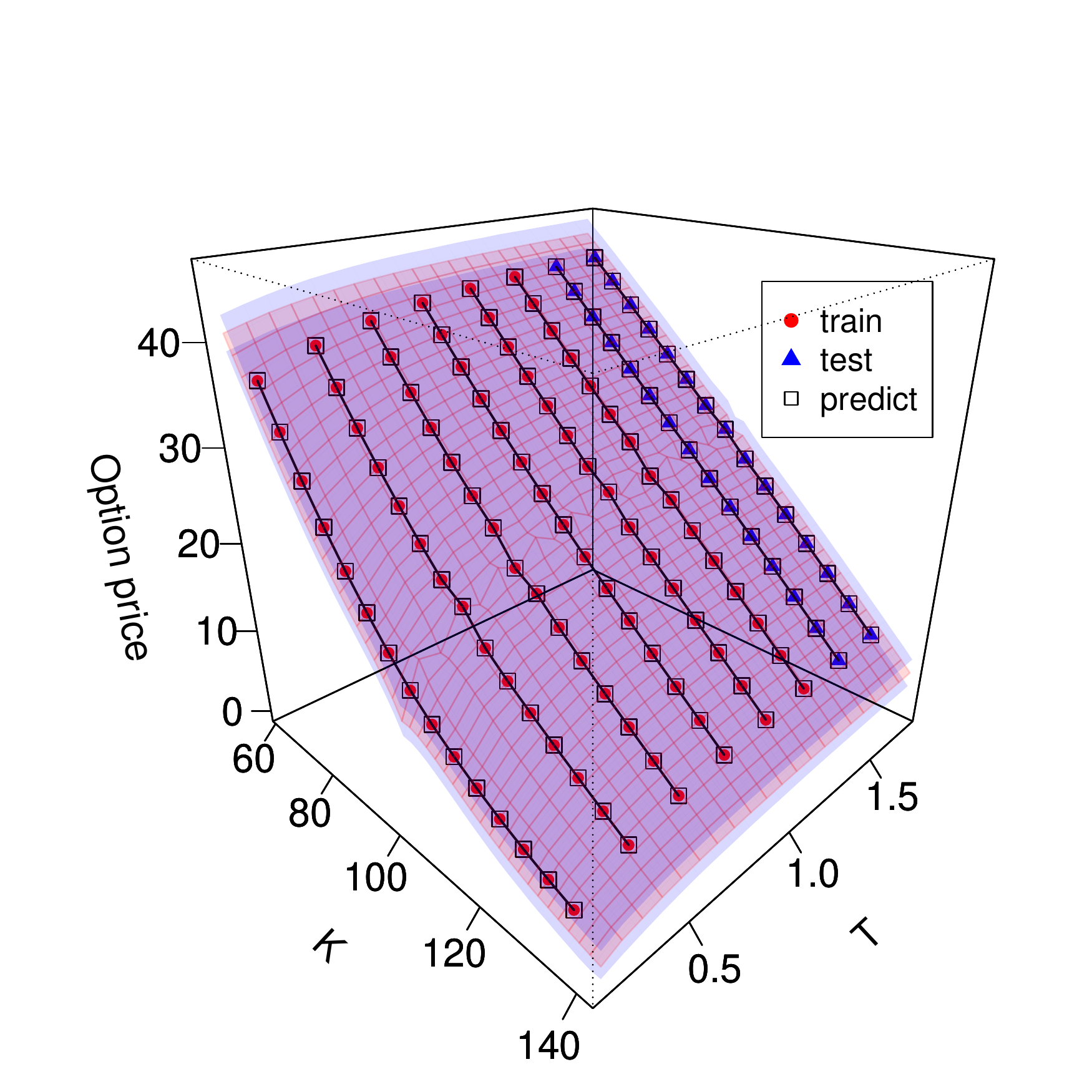}
    \end{center}
    \vspace{-5mm}
    \caption{The red dots are the observed market data for Case 3, the blue triangles are the corresponding held-out data, and the unfilled squares are the corresponding posterior predictive mean. The light blue surfaces are the corresponding 95\% posterior prediction bands.}
    \label{fig::2dlvDataPredcition_kl}
\end{figure}

\begin{figure}[H]
    \begin{center}
       \vspace{-10mm}
        \begin{tabular}[t]{cc}
              \includegraphics[height=8.5cm, width = 8.5cm]{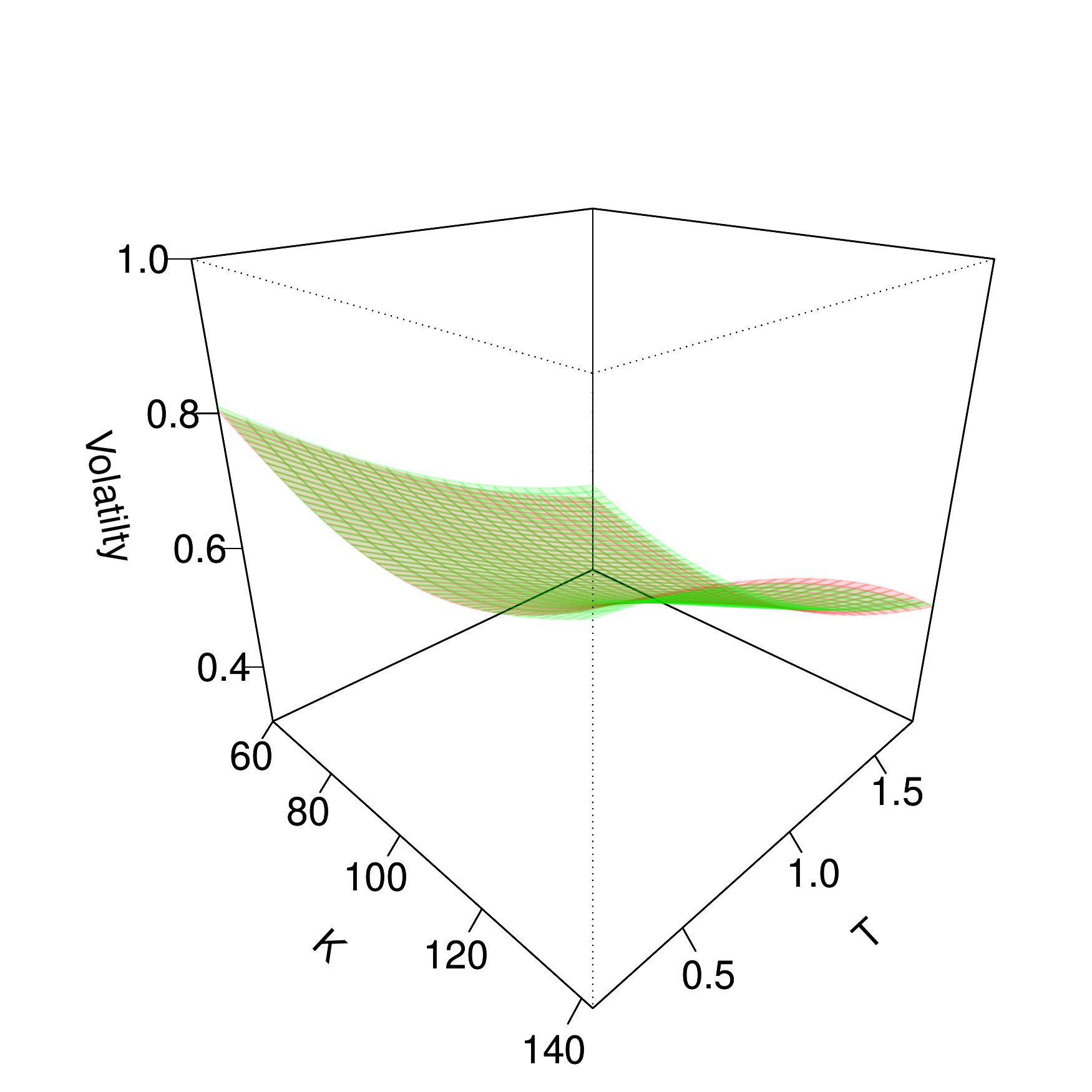}   &  \includegraphics[height=8.5cm, width = 8.5cm]{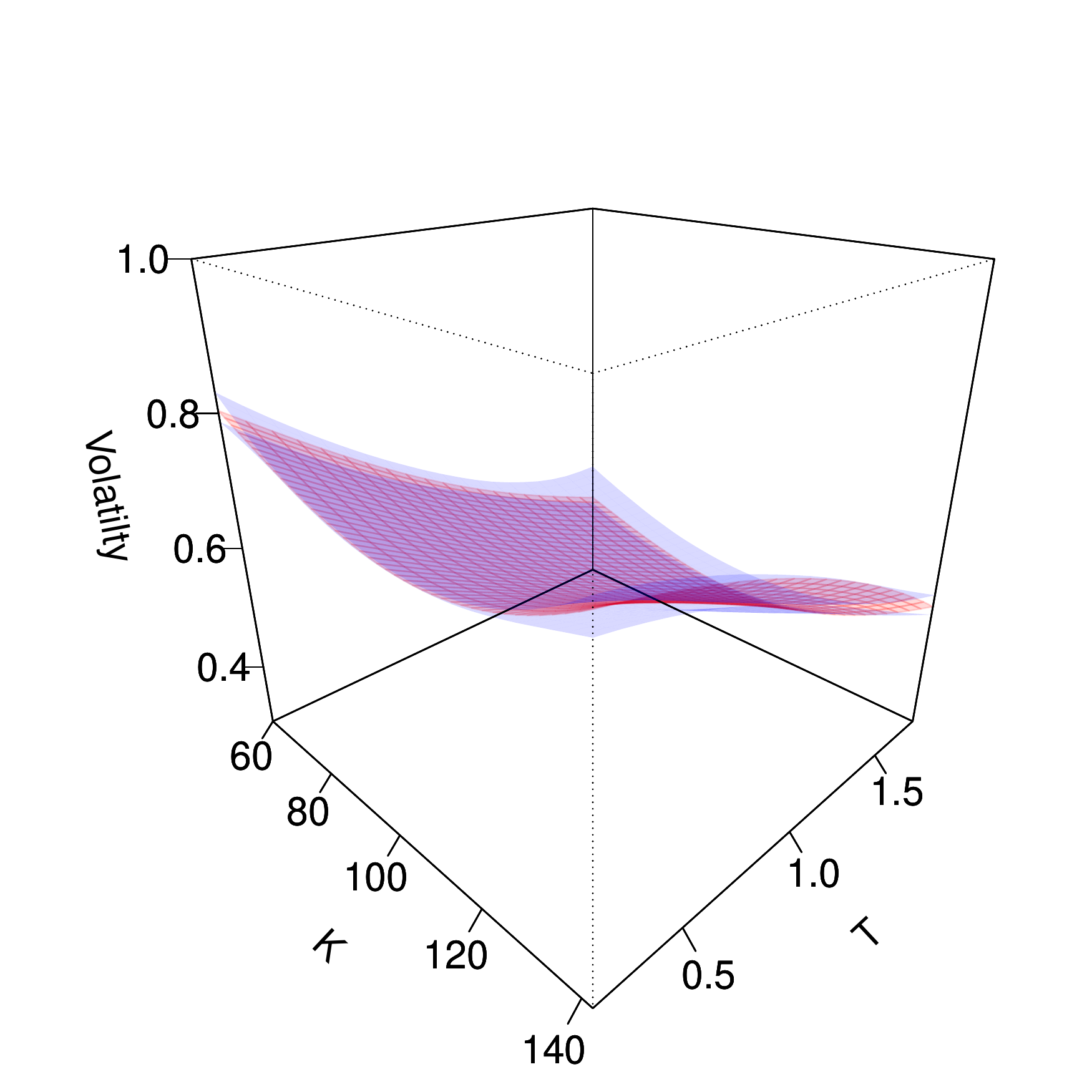}\\
               (a)  &  (b)
        \end{tabular}
    \end{center}
    \vspace{-5mm}
    \caption{(a) The red surface is the benchmark volatility surface for Case 3 and the green surface is the posterior mean; (b) The red surface is the benchmark volatility and the blue surfaces represent the 95\% credible bands.}
    \label{fig::2dlv_kl}
\end{figure}

\begin{figure}[H]
    \begin{center}
    \includegraphics[scale=0.9]{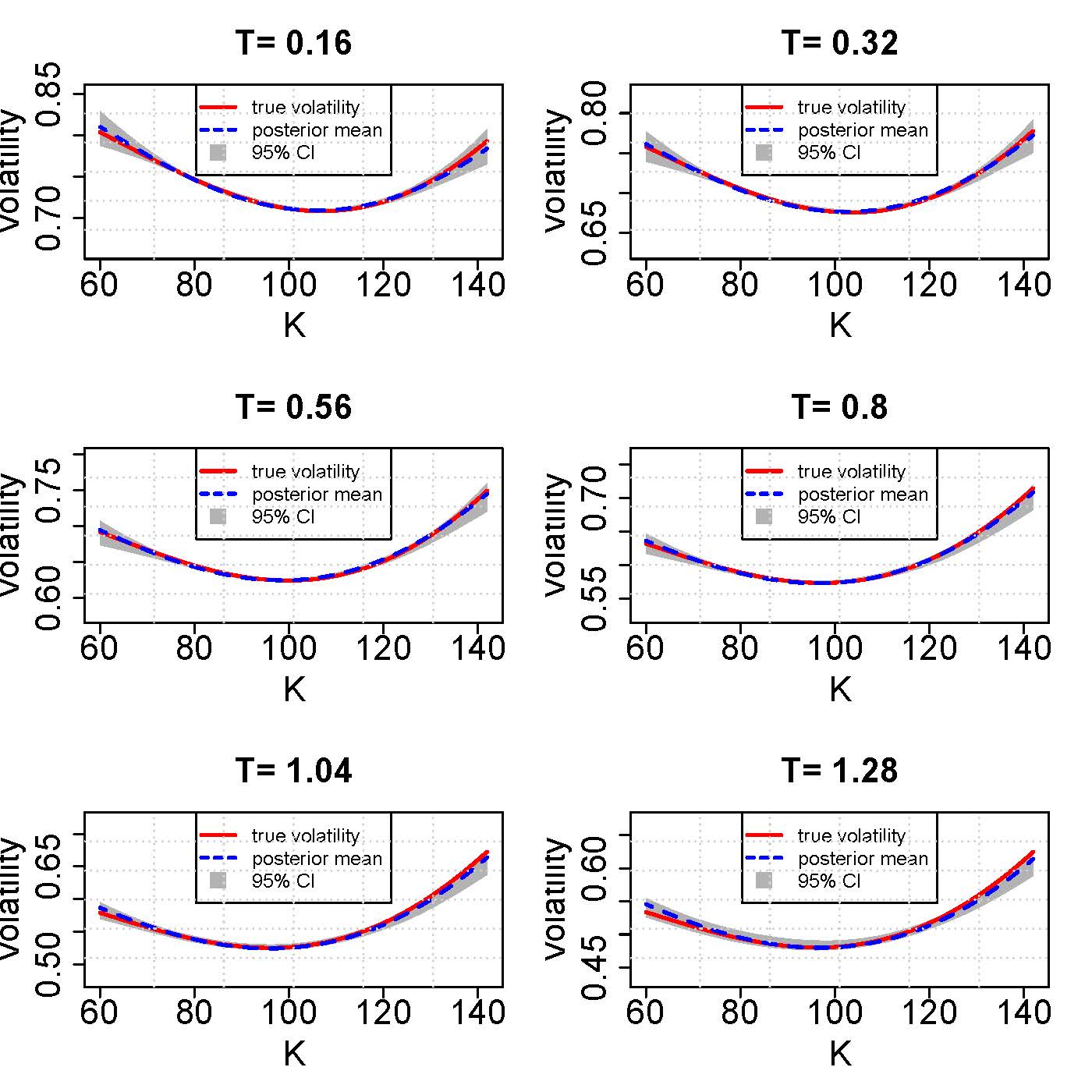}
    \end{center}
    \caption{The slices of local volatility at eight different maturities. The red solid curve is the benchmark volatility in Case 3, the blue dashed curve is the posterior mean, and the grey strip represents the corresponding 95\% credible band.}
    \label{fig::2dlvmaturitybands_case3}
\end{figure}

\begin{figure}[H] 
    \begin{center}
    \includegraphics[scale=0.6]{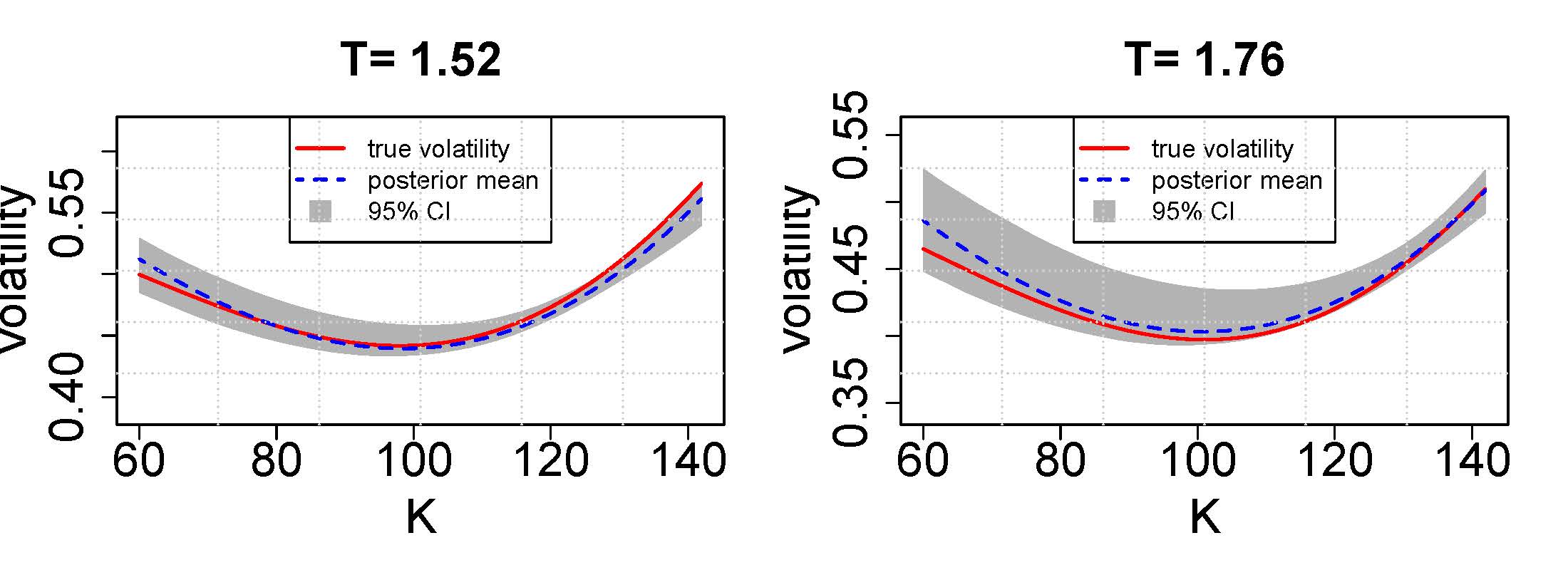}
    \end{center}
    \caption{Posterior predictions for Case 3 for maturity T = 1.52 and T = 1.76. The red curve shows the true volatility, the dashed blue curve shows the posterior predictive mean, and the grey strip represents the 95\% credible bands at different maturities}
    \label{fig::2dlvmaturitybands_prediction_case3}
\end{figure}

In the next synthetic example (Case 3), we generate a reference local volatility surface (transparent blue surface in figure \ref{fig::2dlvDataPredcition_kl}) from a Gaussian process with $l_1 = 0.5$, $l_2 = 0.7$ and $\sigma^2 = 1$  with $T \in [0.25,2]$ and $K \in [65,135]$. A total of 120 option prices are generated. We use options with the first six maturities for calibration and the rest for validation, they are shown as filled red dots and blue triangles respectively in figure \ref{fig::2dlvDataPredcition_kl}. The prior distributions of the parameters are the same as in Case 2. The MCMC iterations, burn-in periods, and thinning frequency are also the same as in Case 2. From figure \ref{fig::2dlv_kl} we can see that our posterior median of volatility surface is very close to the true volatility surface, and the true volatility is contained inside the 95\% credible region. Compared to Case 2, the credible band is much narrower since the data is generated from a Gaussian process, which is the same as the prior model.
Figure \ref{fig::2dlvmaturitybands_case3} represents the snapshots of the volatility surface at different maturity times. From these figures we can see that the posterior mean is very close to true volatility and the 95\% credible bands always contain the true volatility curve. The uncertainties in volatility are lowest around the current price and it increases towards the low and high strike prices. This is consistent with the fact that the market trades heavily on the options with the strikes around the current price. 
Figure \ref{fig::2dlvmaturitybands_prediction_case3} shows the true volatility, the posterior predictive mean, and the corresponding 95\% credible bands at different maturities for the options which has longer maturities but are not traded at the market. We can see that the uncertainties in prediction increase as the maturity time gets longer and the data become sparser. 

\newpage

\subsection{Numerical Results for Market Data} \label{sec::numericalRealData}  

\begin{figure}[H]
    \vspace{-20mm}
    \begin{center}
    \vspace{-10mm}
        \begin{tabular}[t]{cc}
              \includegraphics[height=8.5cm, width = 8.5cm]{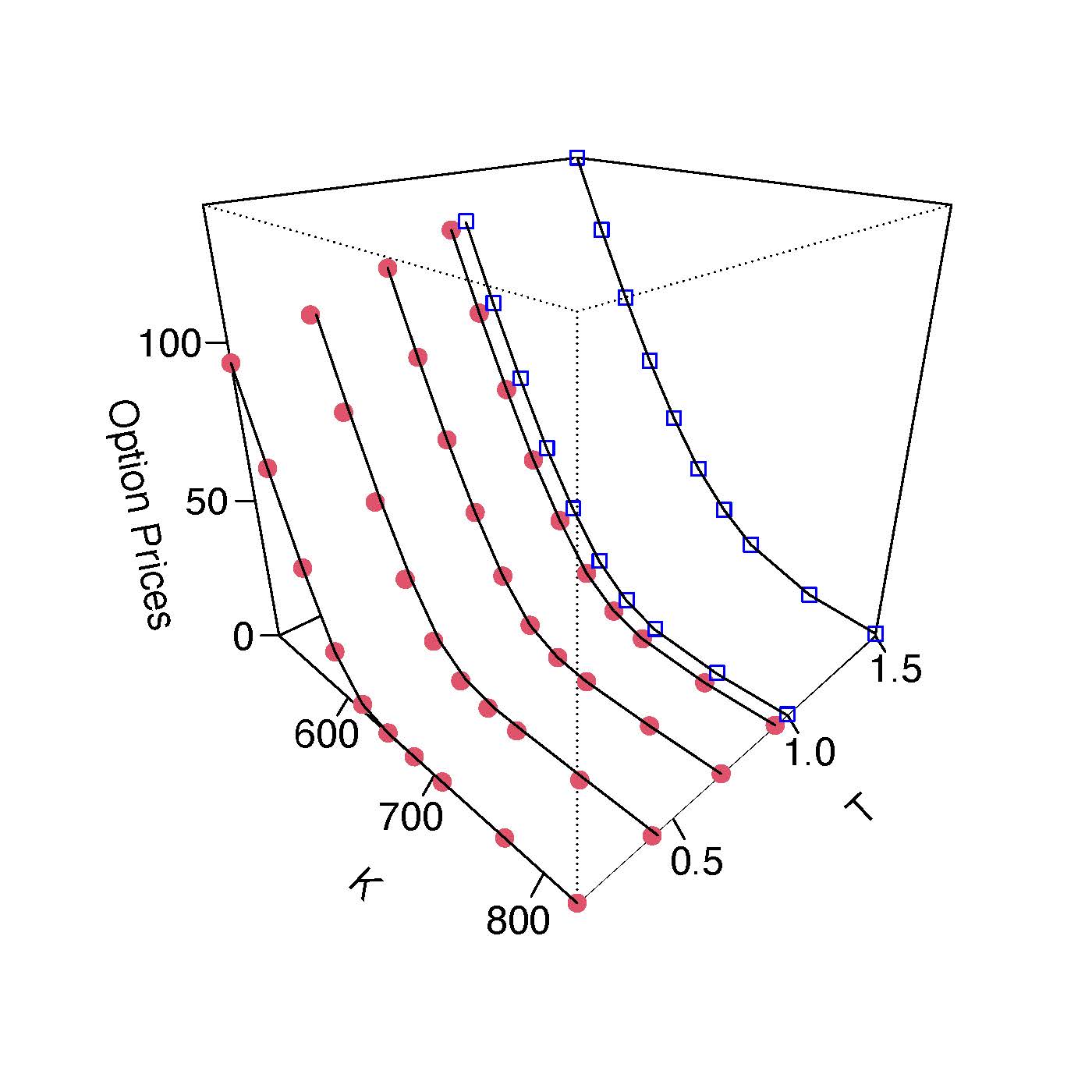}   &  \includegraphics[height=8.5cm, width = 8.5cm]{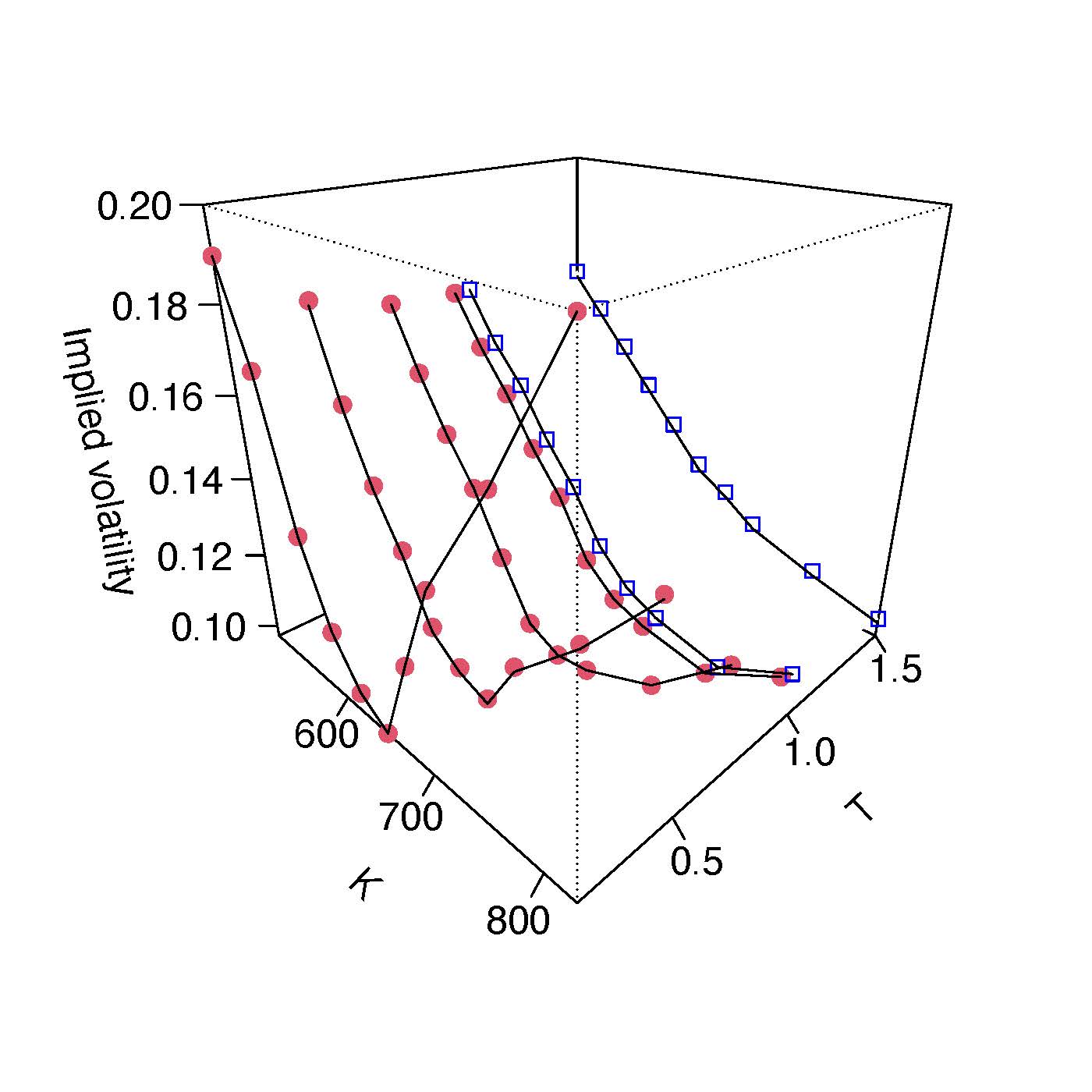}\\
               (a)  &  (b)
        \end{tabular}
    \end{center}
        \caption{(a) S\&P 500 market European option prices in October 1995. Red dots are the data used for local volatility calibration and the blue unfilled rectangles are the data used for validation; (b) The B-S model implied volatility corresponding to (a).}
        \label{fig::real50_data_raw}
\end{figure}

\begin{figure}[H]
    \vspace{0mm}
    \begin{center}
    \vspace{-10mm}
        \begin{tabular}[t]{cc}
              \includegraphics[height=8.5cm, width = 8.5cm]{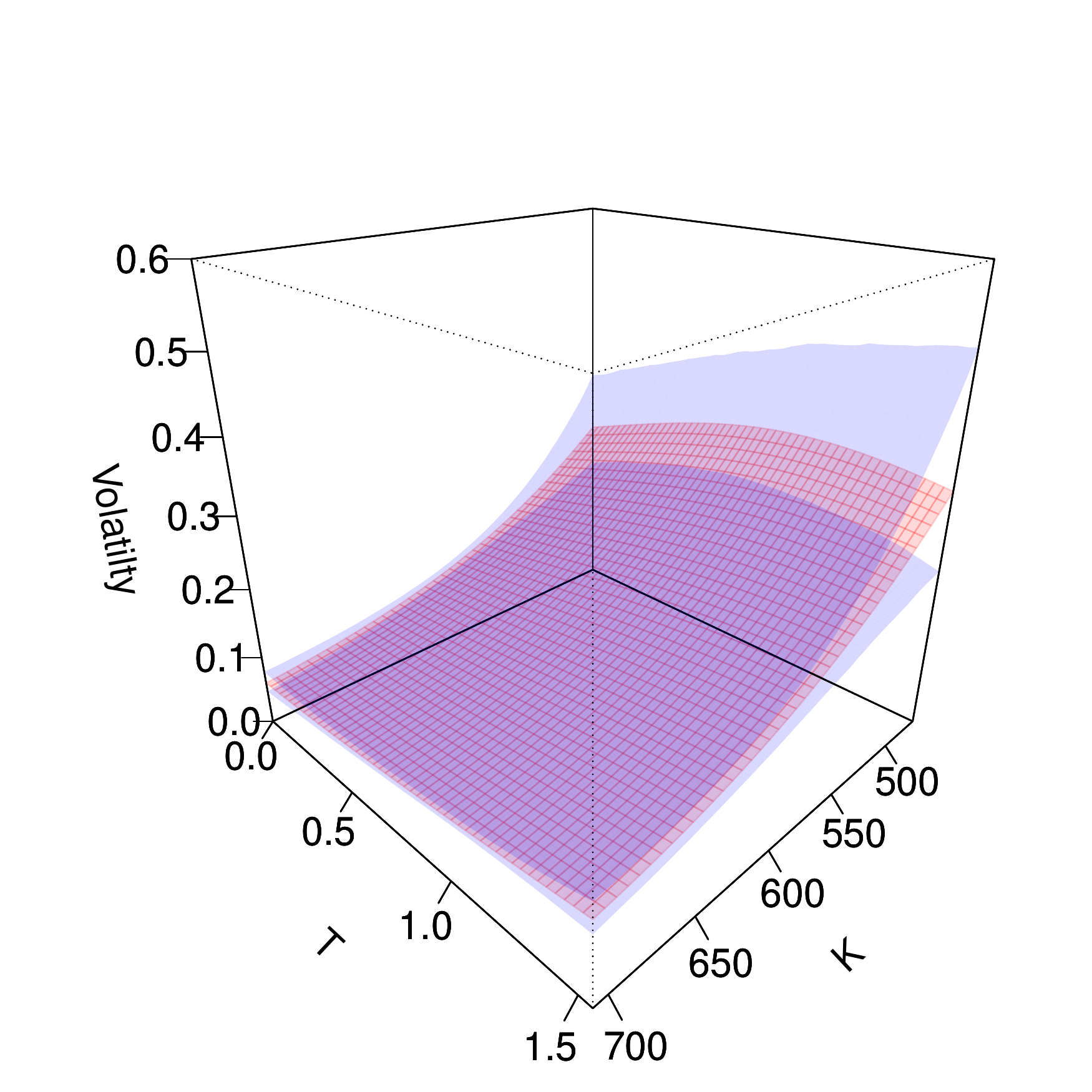}   &  \includegraphics[height=8.5cm, width = 8.5cm]{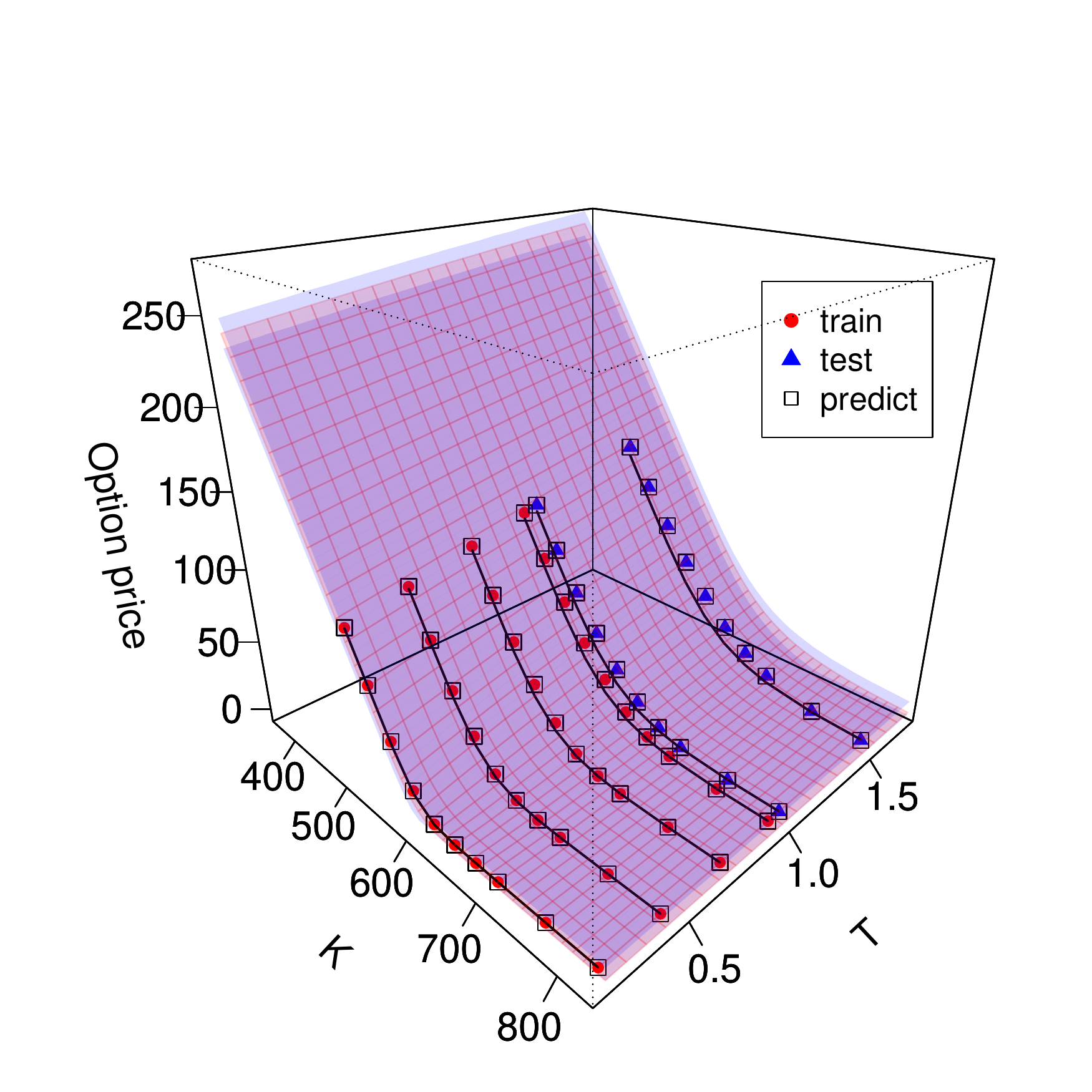}\\
               (a)  &  (b)
        \end{tabular}
    \end{center}
        \caption{(a) The red surface is the posterior mean of volatility, and the blue surfaces are the corresponding 95\% posterior credible bands; (b) The red dots are the observed market data, the blue triangles are corresponding to the held-out data, and the unfilled squares are the corresponding posterior predictive mean. The light blue surfaces are the corresponding 95\% posterior prediction bands.}
        \label{fig::real50_data_fit}
\end{figure}

Next, we apply our methodology in two real market case studies. For the first market case study (Case 4), we use the S\&P 500 equity index European Call option data set containing 57 option prices quoted in October 1995 \citep{Coleman1999ReconstructingTU}. The initial price $S_0$ is $\$590$, interest rate $r$ is $0.06$ and dividend rate $q$ is $0.026$. Figure \ref{fig::real50_data_raw} displays the market option data and its corresponding implied volatility data. Red dots are the observed data used for local volatility estimation and the blue unfilled squares are the held-out data used for validation of the calibrated model. The prior distributions of the parameters are taken to be the same as the synthetic examples in Case 2 and Case 3. We retain $20$ terms in the truncated K-L expansion and run the TSAM sampler $250000$ iterations to sample from the posterior. After $25000$ burn-in we retain every $10$th sample from the chain. Figure \ref{fig::real50_data_fit}(a) shows the posterior mean surfaces and 95\% credible regions from the posterior samples. Figure \ref{fig::real50_data_fit}(b) displays the market option data and the posterior predictive mean of the repriced option price using the estimated volatility surface for both the training and validation data set. The corresponding 95\% prediction bands are also shown on the same plot.

\begin{figure}[H]
    \begin{center}
            \vspace{-20mm}
            \vspace{-10mm}
        \begin{tabular}[t]{cc}
              \includegraphics[height=8.5cm, width = 8.5cm]{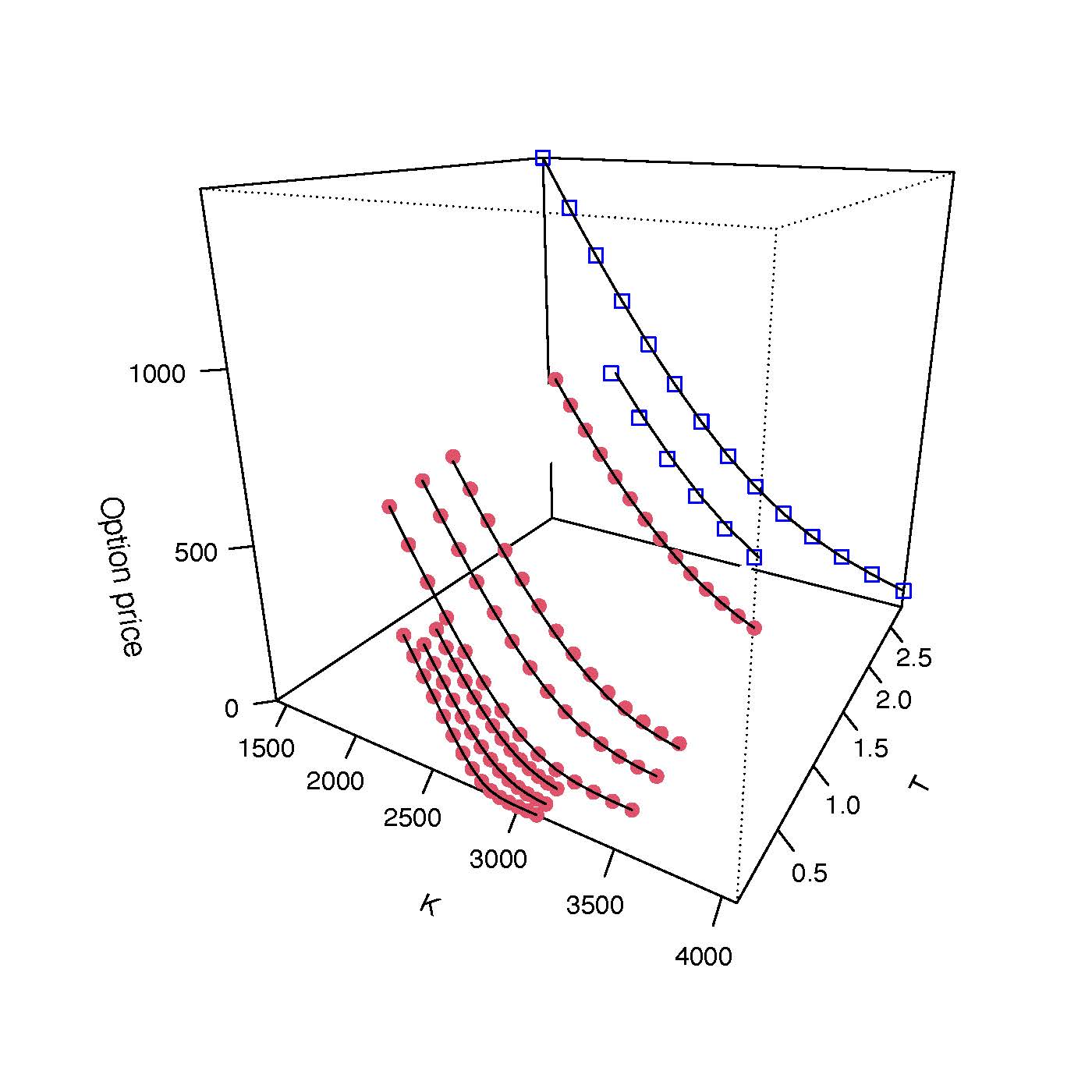}   &  \includegraphics[height=8.5cm, width = 8.5cm]{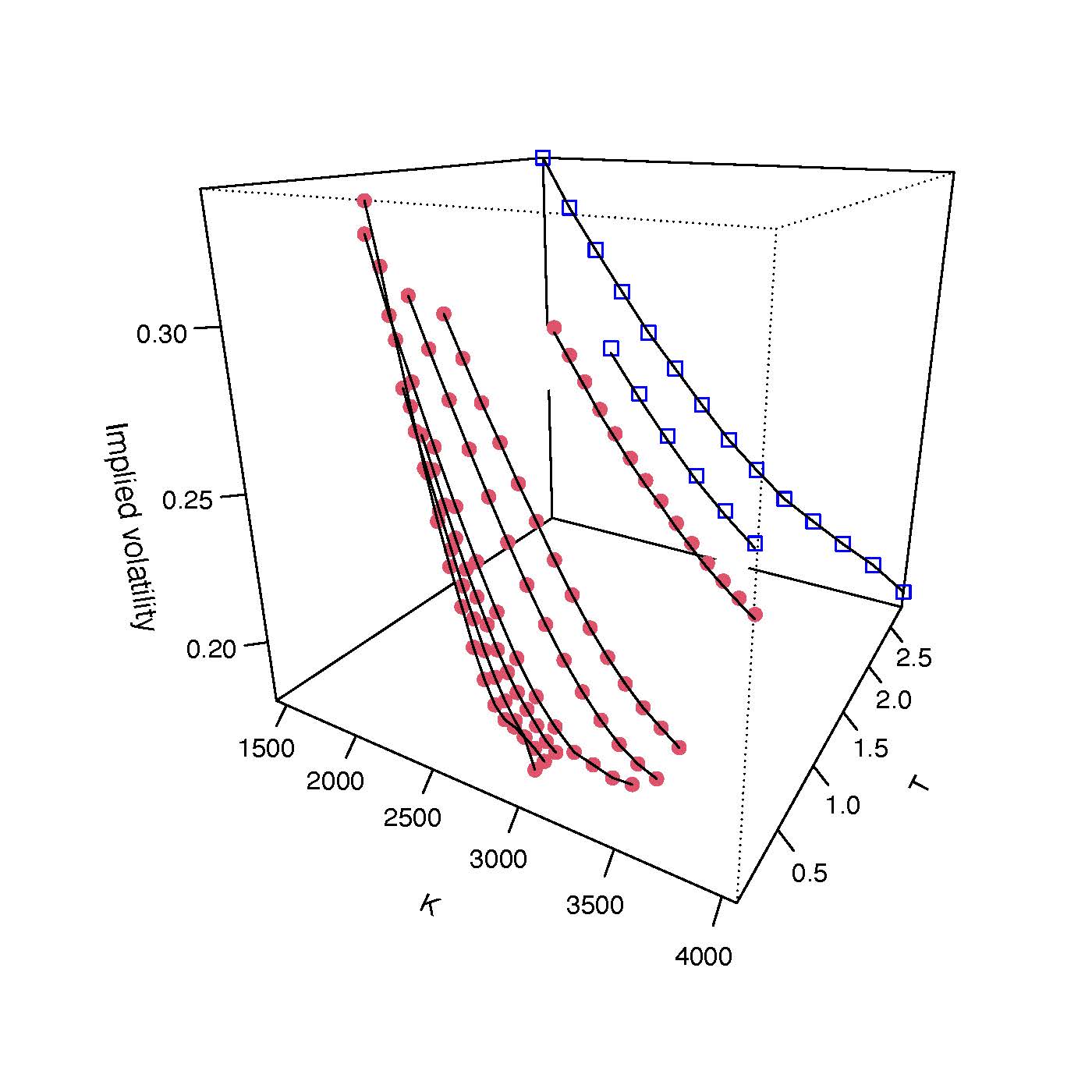}\\
               (a)  &  (b)
        \end{tabular}
    \end{center}
        \caption{(a) Euro Stoxx 50 market European option prices on March 1st 2010. Red solid circles are the data used for local volatility calibration and the blue rectangles are the data used for validation; (b) The B-S model implied volatility corresponding to (a).}
        \label{fig::real155_data_raw}
\end{figure}

\begin{figure}[H]
    \begin{center}
     \vspace{-0mm}
     \vspace{-10mm}
        \begin{tabular}[t]{cc}
              \includegraphics[height=8.5cm, width = 8.5cm]{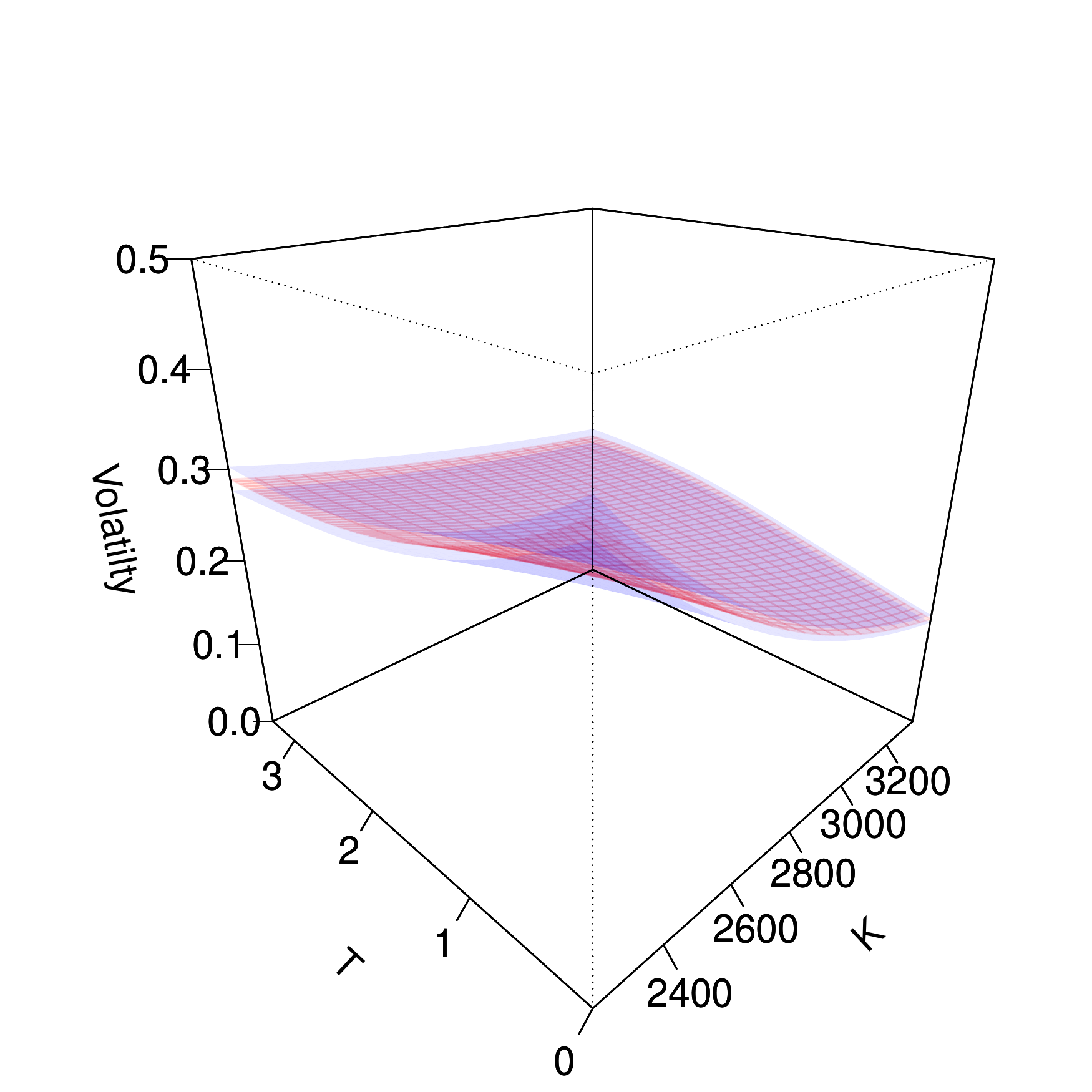}   &  \includegraphics[height=8.5cm, width = 8.5cm]{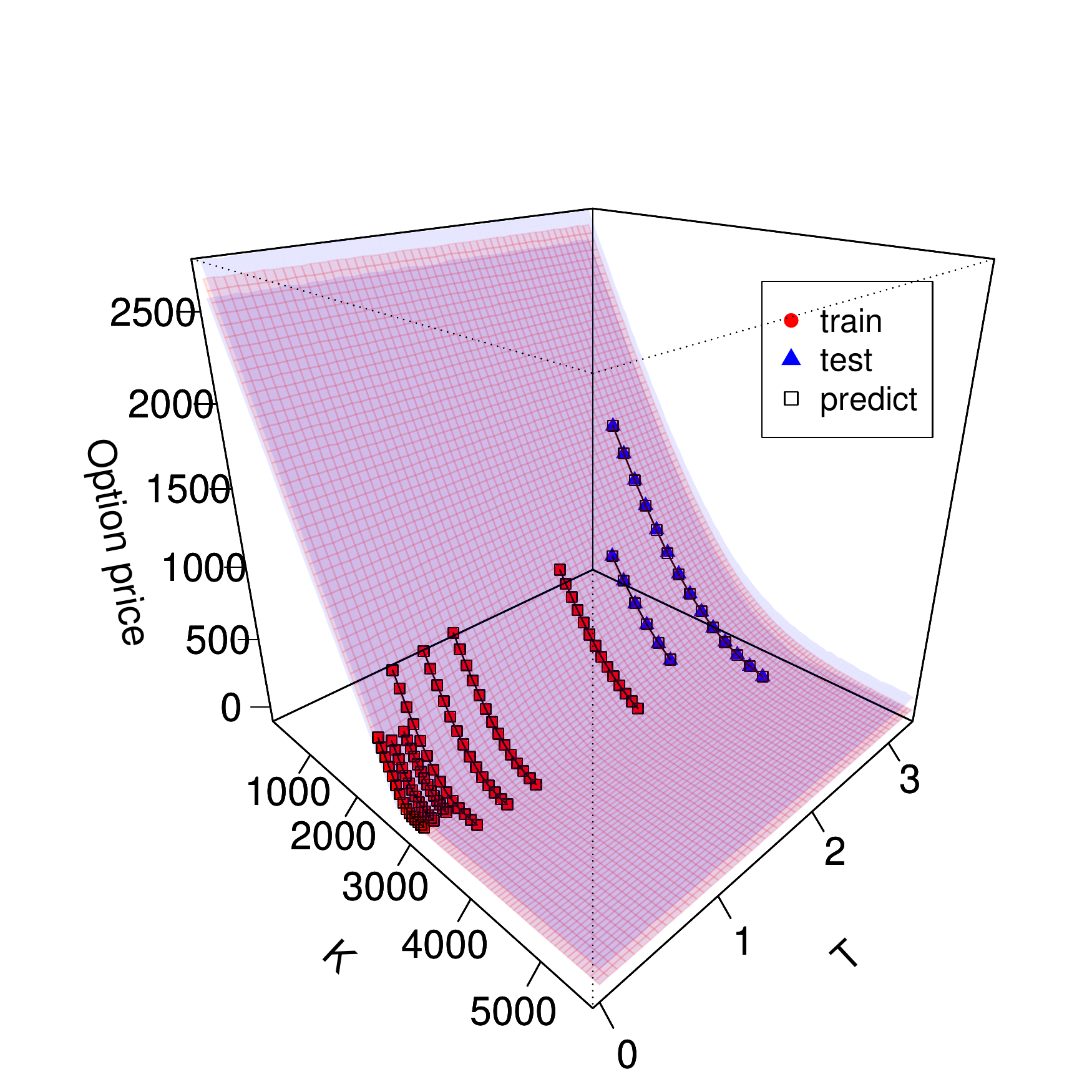}\\
               (a)  &  (b)
        \end{tabular}
    \end{center}
        \caption{(a) The red surface is the posterior mean of volatility, and the blue surfaces are the corresponding 95\% posterior credible bands; (b) The red dots are the observed market data, the blue triangles are corresponding to the held-out data, and the unfilled squares are the corresponding posterior predictive mean. The light blue surfaces are the corresponding 95\% posterior prediction bands.}
         \label{fig::real155_data_fit}
\end{figure}

For the second market case study (Case 5), we use a set of 155 European options on the Euro Stoxx 50 (SX5E) on March 1, 2010, \citep{andreasen2010volatility}. The original market data was in terms of implied volatility. We use the B-S formula to compute the corresponding market option prices with spot price $2772.70$, interest rate $r = 0.01$, and dividend rate $d = 0.034$. Figure \ref{fig::real155_data_raw} displays the market option data and corresponding implied volatility data. The red dots are the observed data used for local volatility estimation and the blue unfilled squares are the held-out data used for validation of the calibrated model. The prior distributions of all the parameters are in the same form as in Cases 2, 3, and 4. We retain $22$ terms in the truncated K-L expansion and run the TSAM sampler $300000$ iterations to sample from the posterior. After $50000$ burn-in we retain every $10$th sample from the chain. Figure \ref{fig::real155_data_fit}(a) shows the posterior mean surfaces and 95\% credible bands for the unknown volatility surface. Figure \ref{fig::real155_data_fit}(b) displays the market option data and the posterior predictive mean of the repriced option price using the estimated volatility surface for both the training and validation data set. The corresponding 95\% prediction bands are also shown on the same plot.

For both the market case studies, similar conclusions to the synthetic data cases can be made. In both the case studies, the posterior predictive mean is very close to the observed option price data for both the training and validation sets, and the 95\% credible bands are also very tight. This demonstrates that our methodology estimates the volatility reasonably well and the estimated volatility can capture market uncertainty in predicting the option prices very accurately. For short maturities, the prices of options with strikes far away from the spot prices are insensitive to volatility,  which justifies the higher volatility uncertainties bands in that region. Another reason for high uncertainties in that region is that low volumes are traded in that area since the chance of profit is relatively low. On the other hand, the option prices at the money are very sensitive to volatility, thus we can see for, across the maturity, the volatility uncertainty near the at-the-money strike is the smallest. This is consistent with other findings in the volatility literature.
For longer maturities, options with strikes far away from spot price become more sensitive to volatility since they have more chance to move in and out of money. On the other hand, the uncertainties in volatility estimates increase around the at-the-money region since the data is sparser for longer maturities. The options in that region will be more likely to move in or out of money with a longer time to expire. In effect, the insensitivity of options over these strikes and maturities makes the likelihood uninformative about local volatility. These facts result in more volatility uncertainties in that region. From numerical experiments of \citep{geng2014non}, the region of less uncertain local volatility surface can be reconstructed from the options with strikes within the interval $[0.8\times S_0, 1.2\times S_0]$. We found similar property such that the uncertainty outside region A in figure\ref{fig::lvSignificant} is large. For example, when T is expiring and $S_0$ is very far from $K$, then the volatility has very high uncertainty, although, in practice, those spots barely have any trading volumes. Another extreme example would be if T is $100$ years, then the time value will dominate and the volatility will have very little influence on the option prices.
\newpage

\newpage

\section{Conclusion}

In this article, we developed a Bayesian methodology to quantify the uncertainty of the unknown local volatility surface given the option data. Our methodology is applied to three synthetic and two real case studies. The calibrated models are able to capture the volatility smile and smirks, and the uncertainties of the estimates provide us a level of confidence about the estimates. The calibrated models are used to reprice and predict European option prices together with the uncertainties. The posterior distribution of local volatility can be of significant value to practitioners who are interested in studying the sensitivity of their hedging strategies to the uncertainty in local volatility value. It also has great importance for pricing exotic options because of the infinitely differentiable property of the Gaussian Process with a squared exponential kernel. The methodology also can be easily adapted to calibration of local volatility from American options, which will be our future research.

In summary, the main contributions of the paper are as follows. First, we use K-L expansion over Gaussian Process (GP) prior which reduces the dimension of the posterior distribution significantly. This makes the MCMC-based posterior inference feasible. In this context, we use the analytical K-L expansion over the GP prior and extend the analytical form of K-L expansion for squared exponential covariance in one dimension to two-dimension using the separable property. The second contribution is the use of the finite element method for solving Dupire's PDE in the Markov chain Monte Carlo setting, which is particularly very important for the observed option data in an irregular grid. The use of the recently developed two-stage adaptive Metropolis for computation efficiency was also another important aspect of our research. Moreover, combining all these modern statistical techniques in a Bayesian framework and developing a methodology to quantify the uncertainties in the local volatility is the main thrust of the paper. For definiteness, here our research is focused on financial models, but the developed methodology can also be used to calibrate complex physical models in other areas of science and engineering which are governed by PDEs and where the unknown PDE coefficients are temporally and/or spatially dependent.

\section*{Conflict of Interest Statement}
The Authors declare that there is no conflict of interest.

\section*{Funding}
This research received no specific grant from any funding agency in the public, commercial, or not-for-profit sectors.

\bibliographystyle{rQUF}
\bibliography{rQUFguide}

\begin{thebibliography}{33}
\providecommand{\natexlab}[1]{#1}
\providecommand{\noopsort}[1]{}
\providecommand{\printfirst}[2]{#1}
\providecommand{\singleletter}[1]{#1}
\providecommand{\switchargs}[2]{#2#1}

\bibitem[\protect\citeauthoryear{Andersen and
  Andreasen}{2000}]{andersen2000jump}
Andersen, L. and Andreasen, J., Jump-diffusion processes: Volatility smile
  fitting and numerical methods for option pricing. {\itshape Review of
  derivatives research}, 2000, \textbf{4}, 231--262.

\bibitem[\protect\citeauthoryear{Andersen and
  Brotherton-Ratcliffe}{1998}]{andersen1998equity}
Andersen, L. and Brotherton-Ratcliffe, R., The equity option volatility smile:
  an implicit finite-difference approach. {\itshape Journal of Computational
  Finance}, 1998, \textbf{1}, 5--37.

\bibitem[\protect\citeauthoryear{Andreasen and
  Huge}{2010}]{andreasen2010volatility}
Andreasen, J. and Huge, B.N., Volatility interpolation. {\itshape Available at
  SSRN 1694972}, 2010.

\bibitem[\protect\citeauthoryear{Animoku
  {\itshape{et~al.}}}{2018}]{AnimokuUY18}
Animoku, A., Ugur, {\"{O}}. and Yolcu{-}Okur, Y., Modeling and implementation
  of local volatility surfaces in Bayesian framework. {\itshape Comput. Manag.
  Sci.}, 2018, \textbf{15}, 239--258.

\bibitem[\protect\citeauthoryear{Avellaneda
  {\itshape{et~al.}}}{1997}]{Avellaneda1997CalibratingVS}
Avellaneda, M., Friedman, C., Holmes, R. and Samperi, D.J., Calibrating
  volatility surfaces via relative-entropy minimization. {\itshape Applied
  Mathematical Finance}, 1997, \textbf{4}, 37--64.

\bibitem[\protect\citeauthoryear{Black and Scholes}{1973}]{black1973pricing}
Black, F. and Scholes, M., The pricing of options and corporate liabilities.
  {\itshape Journal of political economy}, 1973, \textbf{81}, 637--654.

\bibitem[\protect\citeauthoryear{Bollerslev
  {\itshape{et~al.}}}{1988}]{bollerslev1988capital}
Bollerslev, T., Engle, R.F. and Wooldridge, J.M., A capital asset pricing model
  with time-varying covariances. {\itshape Journal of political Economy}, 1988,
  \textbf{96}, 116--131.

\bibitem[\protect\citeauthoryear{Cezaro
  {\itshape{et~al.}}}{2010}]{cezaro2008convex}
Cezaro, A.d., Scherzer, O. and Zubelli, J.P., A convex-regularization framework
  for local-Volatility calibration in derivative markets: the connection with
  convex-risk measures and exponential families. In {\itshape Proceedings of
  the }{\itshape 6th world congress of the bachelier finance society,
  Toronto.}, 2010.

\bibitem[\protect\citeauthoryear{Coleman
  {\itshape{et~al.}}}{1999}]{Coleman1999ReconstructingTU}
Coleman, T., Li, Y. and Verma, A., Reconstructing the Unknown Local Volatility
  Function. {\itshape Journal of Computational Finance}, 1999, \textbf{2},
  77--102.

\bibitem[\protect\citeauthoryear{Crank and
  Nicolson}{1947}]{crank_nicolson_1947}
Crank, J. and Nicolson, P., A practical method for numerical evaluation of
  solutions of partial differential equations of the heat-conduction type.
  {\itshape Mathematical Proceedings of the Cambridge Philosophical Society},
  1947, \textbf{43}, 50â€“67.

\bibitem[\protect\citeauthoryear{Cr{\'e}pey}{2003}]{crepey2003calibration}
Cr{\'e}pey, S., Calibration of the local volatility in a generalized
  black--scholes model using tikhonov regularization. {\itshape SIAM Journal on
  Mathematical Analysis}, 2003, \textbf{34}, 1183--1206.

\bibitem[\protect\citeauthoryear{Dupire}{1994}]{dupire1994pricing}
Dupire, B., Pricing with a smile. {\itshape Risk}, 1994, \textbf{7}, 18--20.

\bibitem[\protect\citeauthoryear{Egger and Engl}{2005}]{Egger2005TikhonovRA}
Egger, H. and Engl, H., Tikhonov regularization applied to the inverse problem
  of option pricing: convergence analysis and rates. {\itshape Inverse
  Problems}, 2005, \textbf{21}, 1027--1045.

\bibitem[\protect\citeauthoryear{Fengler}{2009}]{fengler2009arbitrage}
Fengler, M.R., Arbitrage-free smoothing of the implied volatility surface.
  {\itshape Quantitative Finance}, 2009, \textbf{9}, 417--428.

\bibitem[\protect\citeauthoryear{Geng {\itshape{et~al.}}}{2014}]{geng2014non}
Geng, J., Navon, I.M. and Chen, X., Non-parametric calibration of the local
  volatility surface for European options using a second-order Tikhonov
  regularization. {\itshape Quantitative Finance}, 2014, \textbf{14}, 73--85.

\bibitem[\protect\citeauthoryear{Glaser and Heider}{2012}]{glaser2012arbitrage}
Glaser, J. and Heider, P., Arbitrage-free approximation of call price surfaces
  and input data risk. {\itshape Quantitative Finance}, 2012, \textbf{12},
  61--73.

\bibitem[\protect\citeauthoryear{Hein}{2005}]{hein2005some}
Hein, T., Some analysis of Tikhonov regularization for the inverse problem of
  option pricing in the price-dependent case. {\itshape Zeitschrift f{\"u}r
  Analysis und ihre Anwendungen}, 2005, \textbf{24}, 593--609.

\bibitem[\protect\citeauthoryear{Hirsa}{2012}]{hirsa2012computational}
Hirsa, A., {\itshape Computational methods in finance}, 2012, CRC Press.

\bibitem[\protect\citeauthoryear{Hull and White}{1987}]{Hull1987ThePO}
Hull, J. and White, A., The Pricing of Options on Assets with Stochastic
  Volatilities. {\itshape Journal of Finance}, 1987, \textbf{42}, 281--300.

\bibitem[\protect\citeauthoryear{Jackson
  {\itshape{et~al.}}}{1998}]{jackson1998computation}
Jackson, N., Suli, E. and Howison, S., Computation of deterministic volatility
  surfaces. {\itshape Journal of Computational Finance}, 1998, \textbf{2},
  5--32.

\bibitem[\protect\citeauthoryear{Kahal{\'e}}{2004}]{kahale2004arbitrage}
Kahal{\'e}, N., An arbitrage-free interpolation of volatilities. {\itshape
  Risk}, 2004, \textbf{17}, 102--106.

\bibitem[\protect\citeauthoryear{K{\"o}nig}{2013}]{konig2013eigenvalue}
K{\"o}nig, H., {\itshape Eigenvalue distribution of compact operators},
  Operator Theory: Advances and Applications Vol. 16, , 2013, Birkh{\"a}user.

\bibitem[\protect\citeauthoryear{Lagnado
  {\itshape{et~al.}}}{1997}]{lagnado1997technique}
Lagnado, R., Osher, S. {\itshape et~al.}, A technique for calibrating
  derivative security pricing models: numerical solution of an inverse problem.
  {\itshape Journal of computational finance}, 1997, \textbf{1}, 13--25.

\bibitem[\protect\citeauthoryear{Le~Ma{\^\i}tre and
  Knio}{2010}]{le2010spectral}
Le~Ma{\^\i}tre, O. and Knio, O.M., {\itshape Spectral methods for uncertainty
  quantification: with applications to computational fluid dynamics}, 2010,
  Springer Science \& Business Media.

\bibitem[\protect\citeauthoryear{Lishang and
  Youshan}{2001}]{lishang2001identifying}
Lishang, J. and Youshan, T., Identifying the volatility of underlying assets
  from option prices. {\itshape Inverse problems}, 2001, \textbf{17}, 137.

\bibitem[\protect\citeauthoryear{Loeve}{1977}]{loeve1977elementary}
Loeve, M., Elementary probability theory. In {\itshape Probability theory I},
  pp. 1--52, 1977, Springer.

\bibitem[\protect\citeauthoryear{Mondal
  {\itshape{et~al.}}}{2021}]{Mondal2021ATS}
Mondal, A., Yin, K. and Mandal, A., A Two Stage Adaptive Metropolis Algorithm.
  {\itshape arXiv:2101.00118}, 2021.

\bibitem[\protect\citeauthoryear{Rasmussen}{2003}]{rasmussen2003gaussian}
Rasmussen, C.E., Gaussian processes in machine learning. In {\itshape
  Proceedings of the }{\itshape Summer School on Machine Learning}, pp. 63--71,
  2003.

\bibitem[\protect\citeauthoryear{Roberts and Tegner}{2021}]{Tegner2019APA}
Roberts, S. and Tegner, M., Probabilistic machine learning for local
  volatility. {\itshape Journal of Computational Finance}, 2021, \textbf{25},
  1--50.

\bibitem[\protect\citeauthoryear{Shreve}{2004}]{shreve2004stochastic}
Shreve, S.E., Stochastic calculus for finance II: Continuous-time models. In
  {\itshape Springer Finance}, 2004, Springer Science \& Business Media.

\bibitem[\protect\citeauthoryear{Tikhonov
  {\itshape{et~al.}}}{2013}]{tikhonov2013numerical}
Tikhonov, A.N., Goncharsky, A., Stepanov, V. and Yagola, A.G., {\itshape
  Numerical methods for the solution of ill-posed problems},  Vol. 328, , 2013,
  Springer Science \& Business Media.

\bibitem[\protect\citeauthoryear{Wahba}{1990}]{wahba1990spline}
Wahba, G., {\itshape Spline models for observational data}, 1990, SIAM.

\bibitem[\protect\citeauthoryear{Zhu
  {\itshape{et~al.}}}{1998}]{zhu1997gaussian}
Zhu, H., Williams, C.K., Rohwer, R. and Morciniec, M., Gaussian regression and
  optimal finite dimensional linear models. In {\itshape Neural Networks and
  Machine Learning}, 1998, Springer-Verlag, Berlin.

\end{thebibliography}

\appendix
\section{Proof of Result 1} \label{proof::2deigendecomposition}
Here we prove the fact that for a separable kernel, the multi-dimensional eigenpairs can be obtained from the tensor product of the corresponding one-dimension eigenpairs.

{\bf Result 1}: Suppose $\{\lambda_{1i}, ~\phi_{1i} \}_{i=1,2,3,...}$ are eigen pairs of equation \ref{eq::1dEigenFunGaussianMeasure} along T direction, and $\{\lambda_{2j}, ~\phi_{2j}\}_{j=1,2,3,...}$ are eigen pairs of equation \ref{eq::1dEigenFunGaussianMeasure} along K direction and both $\{\phi_{1i} \}_{i=1,2,3,...}$ and $\{\phi_{2j} \}_{j=1,2,3,...}$ are orthonormal. Then $\{\lambda_{1i} \lambda_{2j}, \phi_{1i} \phi_{2j} \}_{i=1,2,3,...; j=1,2,3,...}$ are eigen paris of equation \ref{eq::twodEigenFun} and $\{\phi_{1i} \phi_{2j} \}_{i=1,2,3,...; j=1,2,3,...}$ are orthonormal. 

\begin{proof}
        First we show that $\{\phi_{1i} \phi_{2j} \}_{i=1,2,3,...; j=1,2,3,...}$ are orthonormal:
        \begin{align} \label{2dsepproof1}
          &{\displaystyle  \int \int \phi_{1i}(T) \phi_{2j}(K) \phi_{1i'}(T) \phi_{2j'}(K) d\mu(T) d\mu(K)} \\
          &\hspace{-6mm}={\displaystyle \int  \phi_{1i}(T) \phi_{1i'}(T) \Big(\int \phi_{2j}(K) \phi_{2j'}(K) d\mu(K)\Big) d\mu(T)} \\
          & \hspace{-6mm} = {\displaystyle  \int  \phi_{1i}(T) \phi_{1i'}(T) \delta_{(j,j')} d\mu(T) }\\
          & \hspace{-6mm} =  {\displaystyle \delta_{(i,i')} \delta_{(j,j')}}  \\
          & \hspace{-6mm} = {\displaystyle \delta_{(\{i,j\},\{i',j'\}})}, \\
        \end{align} 
        where $\delta_{(a,b)}= \begin{cases} 1, ~ if ~ a=b \\  0, ~ if ~ a \neq b. \end{cases}$\\
        Now we plug  in $\phi_{1i}(T) \phi_{2j}(K)$ for $i=1,2,3,...$ and $j=1,2,3,...$ into the left hand side of \ref{eq::twodEigenFun} and obtain:

            \begin{align} \label{2dsepproof2}
           &{\displaystyle\int \int exp\{ -\frac{(T-T')^2}{b_T} \} exp\{ -\frac{(K-K')^2}{b_K} \} \phi_{1i}(T') \phi_{2j}(K') d\mu(T')d\mu(K') }\\
           &\hspace{-6mm}= {\displaystyle \int exp\{ -\frac{(K-K')^2}{b_K} \} \phi_{2j}(K') \Big( \int exp\{ -\frac{(T-T')^2}{b_T} \} \phi_{1i}(T') d\mu(T') \Big) d\mu(K') }\\
           &\hspace{-6mm}= {\displaystyle \int exp\{ -\frac{(K-K')^2}{b_K} \} \phi_{2j}(K') \lambda_{1i}(T)  \phi_{1i}(T) d\mu(K') }\\
           &\hspace{-6mm}= {\displaystyle \lambda_{1i}(T)  \phi_{1i}(T) \int exp\{ -\frac{(K-K')^2}{b_K} \} \phi_{2j}(K') d\mu(K') }\\
         &\hspace{-6mm}= {\displaystyle \lambda_{1i}(T) \phi_{1i}(T)   \phi_{2j}(K) \lambda_{2j}(K)  } \\
         &\hspace{-6mm}= {\displaystyle \lambda_{1i}(T) \lambda_{2j}(K) \phi_{1i}(T)   \phi_{2j}(K)  }
           \end{align} 
           
           This completes the proof that $\{\lambda_{1i} \lambda_{2j}, \phi_{1i}\phi_{2j}\} _{i=1,2,3...:j=1,2,3,...}$ are the eigen-pairs for the corresponding 2-dimensional second Fredholm equation \ref{eq::twodEigenFun}.
\end{proof}

\section{Derivation of the numerical solver}
 First of all, we derive the weak form for (\ref{eq::DupireE}) by taking the inner product with an arbitrary test function v over the domain $\Omega$ of K:
\begin{equation} \label{eq::testfuncprod}
    \int_{\Omega} v{\frac{\partial V}{\partial T} dK -  \int_{\Omega} v \frac{1}{2}\sigma^2(K,T)K^2\frac{\partial^2 V}{\partial K^2}} dK + \int_{\Omega} v rK\frac{\partial{V}}{\partial K} dK =0
\end{equation}
Taking integration by parts and rearranging terms in \ref{eq::testfuncprod}, we get:

\begin{equation}\label{eq::weakform}
    \int_{\Omega} v \frac{\partial V}{\partial T} dK +
    \int_{\Omega} \frac{1}{2}\sigma^2(K,T)K^2 \frac{\partial v}{\partial K} \frac{\partial V}{\partial K} dK + \int_{\Omega} (\frac{\partial (\frac{1}{2}\sigma^2(K,T)K^2)}{\partial K}+rK)v \frac{\partial{V}}{\partial K} dK =0
\end{equation}

Local quadratic functions $U$ are used to approximate the true solution V on the following mesh with Dirichlet boundary conditons:
\begin{equation}
    K_{min} = K_0 < K_1 < \cdots < K_{2M} = K_{max}
\end{equation}
Let $ \xi \in [-1, 1]$, then the quadratic basis and its associated derivatives on each interval are: 
\begin{equation}
    N_1:= -\frac{1}{2}\xi(1-\xi); ~~~ N_2:=1-\xi^2; ~~~ N_3:=\frac{1}{2}\xi(1+\xi).
\end{equation}
\begin{equation}
    N_{1\xi}:= -\frac{1}{2}+\xi; ~~~ N_{2\xi}:=-2\xi^2; ~~~ N_{3\xi}:=\frac{1}{2}+\xi.
\end{equation}
Then, the $i_{th}$ quadratic element can be written as: 
\begin{equation}
    U(K(\xi),T) = U_{2i-2}(T)N_1 + U_{2i-1}(T)N_2 + U_{2i}(T)N_3,
\end{equation}
where $U_{2i-2}$, $U_{2i-2}$ and $U_{2i-2}$ are functions of T only.  
To this end, following the Galerkin Method, we further assume that $U_i$ spans $\{N_1, N_2, N_3\}$. Now, we denote the integral of  (\ref{eq::weakform}) on the $i_{th}$ element $[K_{2i-2}, K_{2i}]$ to be $I_i$, which can be written as: 
\begin{equation}
    I_i = M_i \begin{bmatrix}
    U_{T_{2i-2}}(T) \\
    U_{T_{2i-1}}(T) \\
    U_{T_{2i}}(T)
    \end{bmatrix} + J_i \begin{bmatrix}
        U_{{2i-2}}(T) \\
    U_{{2i-1}}(T) \\
    U_{{2i}}(T)
    \end{bmatrix}
\end{equation}
The $3 \times 3$ mass matrix $M_i$ has entries:
\begin{equation}
    M_{i(p,q)}= \frac{h_i}{2}\int_{-1}^{1} N_pN_qd\xi,
\end{equation}
so that 
\begin{equation}
    M_i=\frac{h_i}{30}\begin{bmatrix}
        4 & 2& -1  \\
        2 & 16 & 2 \\
        -1 & 2 & 4 \\
    \end{bmatrix},
\end{equation}
and the 3x3 stiffness matrix $K_i$ has entries: 
\begin{equation}
J_{i(p,q)}= \int_{-1}^1[ \frac{1}{h_i}\sigma^2(K(\xi), T)K^2N_{p\xi} N_{q\xi} +  (\frac{\partial(\frac{1}{2}\sigma^2K^2)}{\partial K}+rK)N_{p\xi}N_{q\xi}    ]d\xi.
\end{equation}
Thus \ref{eq::weakform} now can be written as a linear system of equations: 
\begin{equation}
\label{K_fin}
    M U_{T} + K U = 0,
\end{equation}
where $U = [U_0(T), U_1(T), \cdots, U_{2M}(T)]^T$ and the pentadiagonal full mass matrix M: 
\begin{equation}\label{eq::fullM}
 M = \frac{1}{30}\begin{bmatrix}
 4h_1 & 2h_1  & -h_1      &       &              &      &      & \cdots & 0       \\
 2h_1 & 16h_1 & 2h_1      & 0     &              &      &      &        & \vdots  \\
 -h_1 & 2h_1  & 4h_1+4h_2 & 2h_2  & -h_2         &      &      &        &         \\
      & 0     & 2h_2      & 16h_2 & 2h_2         & 0    &      &        &         \\
      &       & -h_2      & 2h_2  & 4h_2 + 4 h_3 & 2h_3 & -h_3 &        &         \\
      &       &           & 0     & 2h_3         & 16h_3&\ddots& \ddots &         \\
      &       &           &       & -h_3         &\ddots&\ddots&  2h_M  & - h_M   \\ 
 \vdots&      &           &       &              &\ddots& 2h_M & 16h_M  &  2h_M   \\
    0  & \cdots&          &       &              &      & -h_M &  2h_M  &  4h_M   \\
\end{bmatrix}
\end{equation}

By in-cooperating the Dirichlet boundary conditions, we get: 
\begin{equation} \label{eq::spatialsystem}
    MU_T + JU = b^*
\end{equation}
where $U = [U_1(T), U_2(T), \cdots, U_{2M-1}(T)]^T$ and 
\begin{equation}
    b^* = \begin{bmatrix}
    -2h_1u_{a_T}(T) - J_{1,0} u_a(T) \\
    h_1u_{a_T}(T) - J_{2,0} u_a(T) \\
    0 \\
    \vdots \\
    0 \\
    h_Mu_{b_T}(T) - J_{2M-2,2M} u_b(T) \\
    -2h_Mu_{b_T}(T) - J_{2M-1,0} u_b(T) \\
    \end{bmatrix}
\end{equation},

$u_a(T)$ and $u_b(T)$ are the left and right Dirichlet boundary conditions not dependent on K. 
This completes the finite-element step along $K$ direction keeping $T$ fixed.
We now apply the Crank-Nicolson type finite difference method \citep{crank_nicolson_1947} to the dynamic system odes \ref{eq::spatialsystem}. 
We discretize the temporal dimension as:
\begin{equation}
0 = T_0 < T_1 < \cdots < T_n < \cdots < T_N = T_{max},
\end{equation}
and define $k_n := T_n - T_{n-1}$.

We solve (\ref{eq::spatialsystem}) with forward Euler and backward Euler, and take the average of the two solutions as:
\begin{equation}
    M \frac{(U^n - U^{n-1})}{\frac{1}{2}k_n} + K U^{n} = b^n
\end{equation}

\begin{equation}
    M \frac{(U^n - U^{n-1})}{\frac{1}{2}k_n} + K U^{n-1} = b^{n-1}
\end{equation}

\begin{equation}
\label{T_fin}
    (M+\frac{1}{2}k_nK^n)U^n = (M-\frac{1}{2}k_nK^{n-1})U^{n-1}+b
\end{equation}
where

\begin{equation}
    b= \begin{bmatrix}
     -2h_1(u_a^n-u_a^{n-1}) - \frac{1}{2}k_n(K_{1,0}^nu_a^n+K_{1,0}^{n-1}u_a^{n-1}\\
     h_1(u_a^n-u_a^{n-1}) - \frac{1}{2}k_n(K_{2,0}^nu_a^n+K_{2,0}^{n-1}u_a^{n-1}\\
     0\\
 \vdots\\
     0\\
     h_M(u_b^n-u_b^{n-1}) - \frac{1}{2}k_n(K_{2M-2,2M}^nu_b^n+K_{2M-2,2M}^{n-1}u_b^{n-1}\\
     -2h_M(u_b^n-u_b^{n-1}) - \frac{1}{2}k_n(K_{2M-1,2M}^nu_b^n+K_{2M-1,2M}^{n-1}u_b^{n-1}\\
    \end{bmatrix}
\end{equation}
and 
\begin{equation}
    U^0 = [U_{0_1}, U_{0_2}, \cdots, U_{0_{2M-1}}]^T
\end{equation}

Finally we solve the linear systems \ref{K_fin} and \ref{T_fin} iteratively to obtain the solution for \ref{eq::DupireE}.
\end{document}